\documentclass[fleqn,usenatbib]{mnras}

\usepackage{newtxtext,newtxmath}

\usepackage[T1]{fontenc}

\DeclareRobustCommand{\VAN}[3]{#2}
\let\VANthebibliography\thebibliography
\def\thebibliography{\DeclareRobustCommand{\VAN}[3]{##3}\VANthebibliography}

\usepackage{graphicx}	
\usepackage{amsmath}	

\title[Bar likelihood in local galaxies]{Bar likelihood in local galaxies: the joint role of stellar mass, morphology and environment}

\author[M. A. Chac\'{o}n, B. Cervantes Sodi, Y. Jaff\'{e}]{
Manuel Alejandro Chac\'{o}n,$^{1}$\thanks{E-mail: manuel.chacon@utp.ac.pa (MACH)},
Bernardo Cervantes Sodi,$^{2}$\thanks{E-mail: b.cervantes@irya.unam.mx (BCS)},
Yara Jaff\'{e}$^{3,4}$
\\
$^{1}$Direcci\'{o}n Nacional de Ciencias Espaciales, Universidad Tecnol\'{o}gica de Panam\'{a}, Llano Mar\'{i}n, V\'{i}a Panamericana, Provincia de Cocl\'{e}, Panam\'{a}\\
$^{2}$Instituto de Radioastronom\'{i}a y Astrof\'{i}sica, Universidad Nacional Aut\'{o}noma de M\'{e}xico, Antigua Carretera a P\'{a}tzcuaro \# 8701, \\ Ex-Hda. San Jos\'{e} de la Huerta, Morelia, Michoac\'{a}n, C.P. 58089, M\'{e}xico\\
$^{3}$ Departamento de F\'{i}sica, Universidad T\'{e}cnica Federico Santa Mar\'{i}a, Avenida Espa\~{n}a 1680, Valpara\'{i}so, Chile\\
$^{4}$Millennium Nucleus for Galaxies (MINGAL)
}

\date{Accepted XXX. Received YYY; in original form ZZZ}

\pubyear{\the\year{}}

\begin{document}
\label{firstpage}
\pagerange{\pageref{firstpage}--\pageref{lastpage}}
\maketitle

\begin{abstract}
We investigate how the likelihood of hosting a stellar bar depends on both intrinsic galaxy properties and environment, using a volume-limited sample of central and satellite galaxies from the NASA--Sloan Atlas, spanning isolated systems, low- and high-richness groups, and massive haloes, complemented with halo-based environmental information and probabilistic morphology. Using probabilistic bar identification, we quantify the bar likelihood as a function of stellar mass, colour, morphology, halo mass, and cluster-centric distance. We find that stellar mass and morphology are the galaxy properties most associated with the bar likelihood. At fixed morphology, the probability of hosting a bar increases with stellar mass, particularly for late-type galaxies. At the same time, morphology strongly modulates the overall incidence of bars, with early-type galaxies displaying systematically lower probabilities than late-type systems across the stellar mass range. Environmental trends are observed, including halo mass and cluster-centric distance, but these trends largely reflect the morphological composition of the galaxy population. In particular, the decrease in bar likelihood towards the centres of massive haloes reflects the increasing fraction of early-type galaxies, consistent with the morphology--density relation, rather than a direct environmental suppression of bars. Our residual analysis finds no statistically significant residual dependence on halo mass or normalized cluster-centric distance after accounting for stellar mass and morphology, within the sensitivity of this sample. Overall, our results show that stellar mass and morphology are the main factors associated with bar likelihood, while environmental trends largely reflect morphological composition.
\end{abstract}




\section{Introduction}

Bars are among the most prominent non-axisymmetric structures observed in disc galaxies, and are observed in a large fraction of both spiral and lenticular systems \citep{Eskridge+2000, Nair+2010b, Masters+2011, Lee+2019}. Their formation is commonly attributed to dynamical instabilities that drive stellar orbits away from circular symmetry \citep{Athanassoula+1983, Contopoulos+1989}. These structures play a fundamental role in the internal dynamics of galaxies, as they efficiently redistribute the angular momentum among stars, gas, and dark matter, thereby reshaping their internal structure \citep{Weinberg+1985, Sellwood+2014, Gadotti+2020}. Bars can channel gas into the central regions, enhancing nuclear star formation and influencing bulge growth, while also contributing to the kinematic heating of the disk \citep{Combes+1990, Ellison+2011, Seo+2019}. This gas inflow has also motivated studies of a possible link between bars
and nuclear activity, with some works reporting an enhanced incidence of
AGN in barred galaxies \citep{Alonso+2013, Marels+2025} and others finding
no such enhancement once host-galaxy properties are controlled for
\citep{Lee+2012}, suggesting that the connection depends on sample selection
and AGN definition. Furthermore, their frequency and properties appear to depend on galaxy parameters such as stellar mass, colour, and morphology, with evidence suggesting an evolutionary trend in the bar fraction with redshift \citep{Sheth+2008, Melvin+2014}.

An important open question in galaxy evolution concerns the physical conditions that allow a disc galaxy to develop and sustain a stellar bar. A common observational approach is to study bar fractions, defined as the ratio of barred to disc galaxies as a function of various galactic properties. Among these properties, stellar mass has been identified as a key parameter, with observational evidence showing a correlation between bar fraction and stellar mass \citep{Diaz-Garcia+2016}. In addition to stellar mass, several other galaxy characteristics—such as gas content, dark matter halo properties, angular-momentum retention, and bulge structure—also play a significant role in determining the likelihood of bar formation \citep{Lee+2012, Romeo+2023}. In particular, the amount of gas within a galaxy has a strong impact on both the initial development of a stellar bar and its subsequent evolution \citep{Masters+2012,Cervantes-Sodi+2017,Mukundan+2025}.

The role of environment in the formation/dissolution and evolution of stellar bars remains a matter of ongoing debate, with observational studies yielding apparent conflicting results, particularly in dense environments such as galaxy clusters. Early works reported an enhanced fraction of barred galaxies in high-density regions, including the cores of clusters such as Coma, as well as an increased incidence of barred early-type spirals in dense environments \citep{Thompson+1981,Giuricin+1993}. Subsequent studies of both local and intermediate-redshift clusters broadly supported this picture, while also suggesting a more nuanced dependence on galaxy properties, such as luminosity, with interactions potentially triggering bar formation in massive, dynamically stable discs but suppressing it in lower-mass systems through disk heating \citep{Eskridge+2000, Barazza+2009, Mendez-Abreu+2012}. In contrast, a large number of studies have reported little to no dependence of the bar fraction on environment when comparing cluster and field populations \citep{van_den_Bergh+2002, Marinova+2009,Aguerri+2009,Mendez-Abreu+2010, Giordano+2011, Marinova+2012}, although secondary trends have been identified in specific subsamples, such as an increased bar fraction in lenticular galaxies toward cluster centres  \citep{Barway+2011, Lansbury+2014}. Clustering-based analyses have further complicated this picture: studies based on large-scale statistics have found that, at fixed stellar mass, barred and unbarred galaxies exhibit similar clustering properties and thus no clear environmental dependence \citep{Li+2009}, although a later study by \citet{Lin+2014} reported a residual dependence when separating by morphology, with early-type barred galaxies being more strongly clustered than their unbarred counterparts, while late-type systems show the opposite trend. At smaller scales, interactions with close companions appear to play a suppressive role, with the bar fraction decreasing at small separations, possibly due to strong tidal effects that inhibit bar formation or growth \citep{Lee+2012,Casteels+2013, Li2026}.

More recent studies reinforce this complex scenario. While some works argue
that bar formation is largely governed by internal processes, with little or
no dependence on environment once galaxy properties such as stellar mass and
colour are controlled for \citep{Lee+2012, Lin+2014, Sarkar+2021}, others
highlight that environmental effects may still manifest through morphology-
or scale-dependent trends. In this context, \citet{Tawfeek+2022}, using a
sample of galaxies in 32 nearby X-ray selected clusters from the OmegaWINGS
survey, found that the bar fraction strongly depends on stellar mass and
morphology, while the apparent dependence on cluster mass or cluster-centric
distance largely vanishes once morphological type is accounted for.
Complementary evidence has been provided by recent structural studies, which
suggest that the impact of environment extends beyond bar occurrence to their
physical properties. In particular, \citet{Cuomo+2026} find that bars in
cluster environments are systematically shorter and less prominent than
those in filaments and the field, consistent with a scenario in which
environmental processes such as gas stripping, strangulation, and tidal
interactions slow down the secular evolution of bars in high-density regions.

However, much of the observational literature has been framed in
terms of cluster--field comparisons or selected cluster samples, making it
difficult to follow environmental trends continuously across a broad range
of halo masses and to separate the roles of central and satellite galaxies.
In this work, we extend this picture by using a halo-based group catalogue
that allows us to probe a broad range of environments, from isolated systems
through low- and high-richness groups to massive haloes. We analyse central
and satellite galaxies separately and adopt a probabilistic estimate of bar
likelihood rather than a binary barred/unbarred classification. Most
importantly, we quantify the residual environmental dependence of bar
likelihood after accounting simultaneously for stellar mass and morphology.
This combination allows us to test whether the environmental trends seen in
the global bar population persist once the changing intrinsic properties and
morphological composition of galaxies are controlled for.

In this paper, we investigate how the probability of hosting a bar depends
on galaxy properties and environment. Using a well-defined sample of central
and satellite galaxies, we examine: (i) the relation between bar probability
and stellar mass; (ii) the joint distribution of stellar mass,
morphology, and halo mass for satellite systems; (iii) the dependence of bar
likelihood on halo mass for central and satellite galaxies; and (iv) the
variation of bar probability as a function of normalized cluster-centric
distance. We then quantify the residual environmental dependence after
accounting for stellar mass and morphology. This approach allows us to
distinguish population-driven environmental trends from residual variations
in bar likelihood associated with halo-scale environment.
Throughout this work, we adopt a flat $\Lambda$CDM cosmology with 
$H_{0}=70~\mathrm{km~s^{-1}~Mpc^{-1}}$, $\Omega_{\mathrm{m}}=0.3$, and 
$\Omega_{\Lambda}=0.7$.

\section{Observational Data and Methodology}\label{sec:data}

\subsection{Sample Selection}\label{subsec:sample}

Our parent catalogue is the NASA--Sloan Atlas (NSA; \citealt{NSA}),
a reprocessed version of the Sloan Digital Sky Survey (SDSS) designed
to provide improved photometric measurements for nearby galaxies.
In this work, we make use of the spectroscopic redshifts and the
absolute magnitudes derived from S\'ersic profile fits provided by the NSA.

We construct a volume-limited sample using the spectroscopic redshift ($z$) 
and the absolute magnitude in the $r$ band, $M_r$.
The redshift range is restricted to $0.02 < z < 0.055$, which minimizes the
impact of peculiar velocities at very low redshift while maintaining high
spectroscopic completeness over the selected volume.
An absolute magnitude cut of $M_r < -18.5$ is applied to exclude
low-luminosity dwarf galaxies. This limit ensures that the sample remains complete within the adopted redshift range and focuses on
galaxies for which structural features, such as stellar bars, can be
reliably identified. The combination of these redshift and magnitude cuts yields a
well-defined volume-limited sample, minimizing selection effects
such as Malmquist bias and enabling a consistent comparison with
previous studies adopting similar selection criteria.

After applying these selection criteria, the resulting parent sample comprises 39,720 galaxies.

\subsection{Environmental and Physical Properties}\label{subsec:properties}

\begin{figure*}
    \centering
    \includegraphics[width=0.32\textwidth]{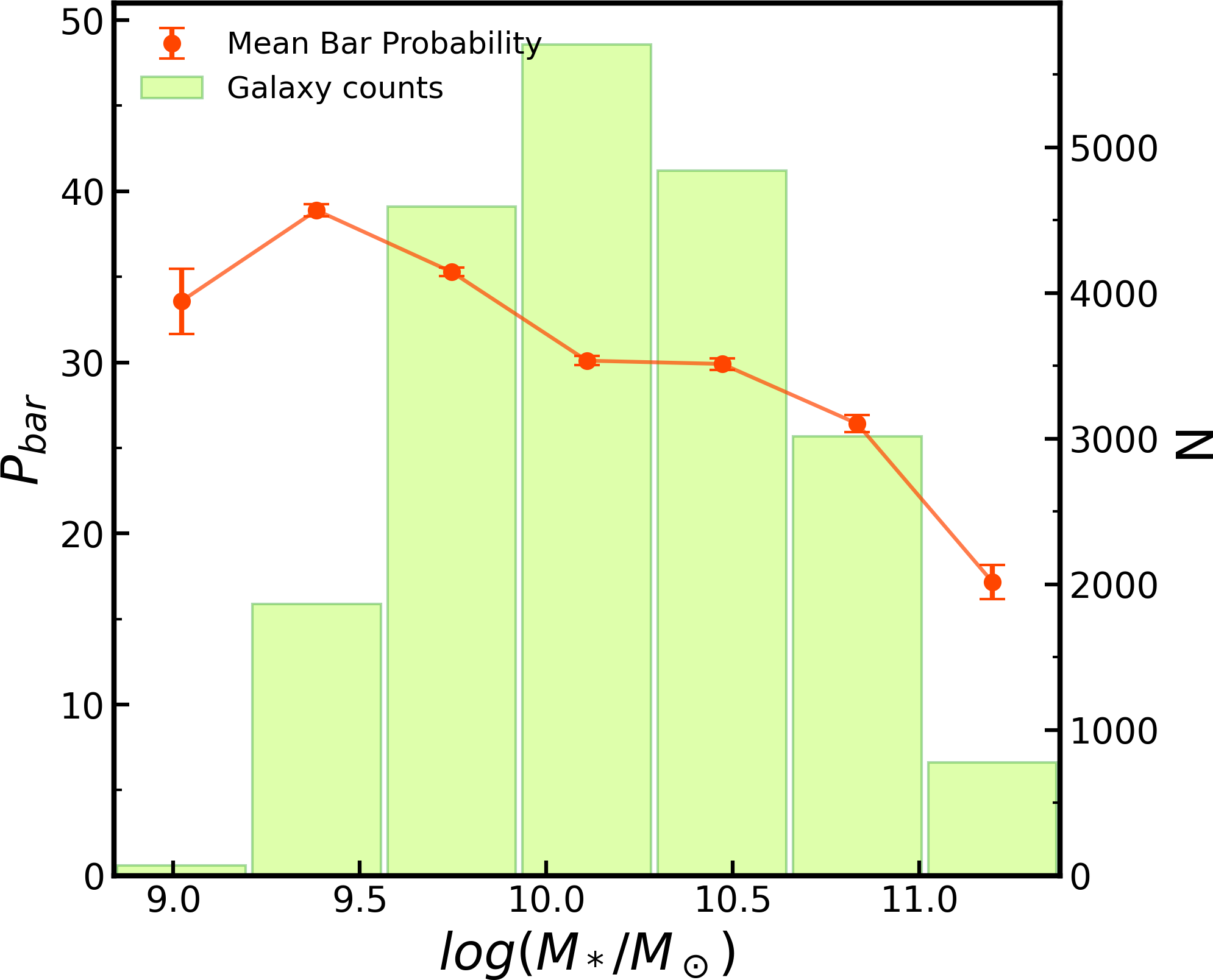}\hfill
    \includegraphics[width=0.32\textwidth]{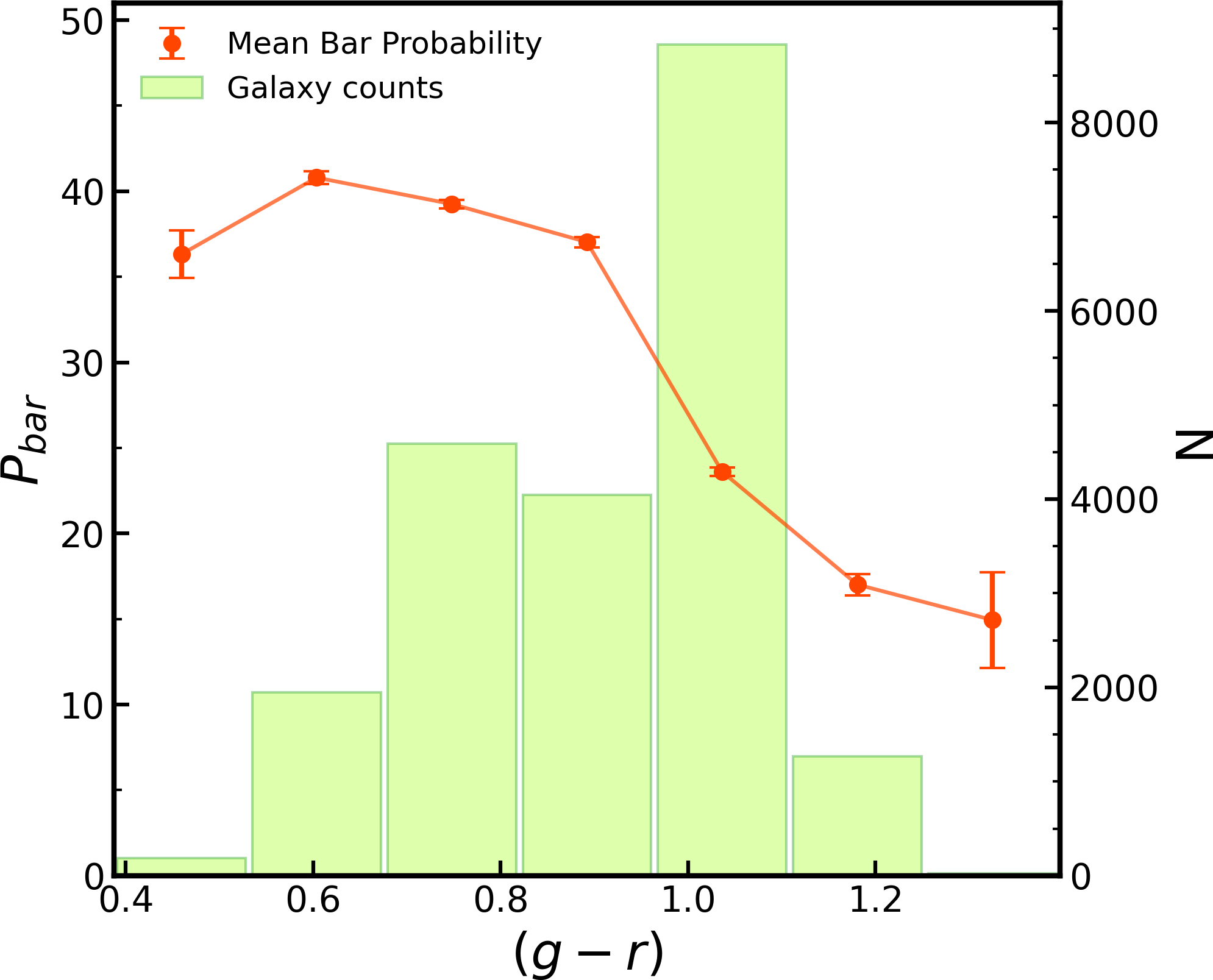}\hfill
    \includegraphics[width=0.32\textwidth]{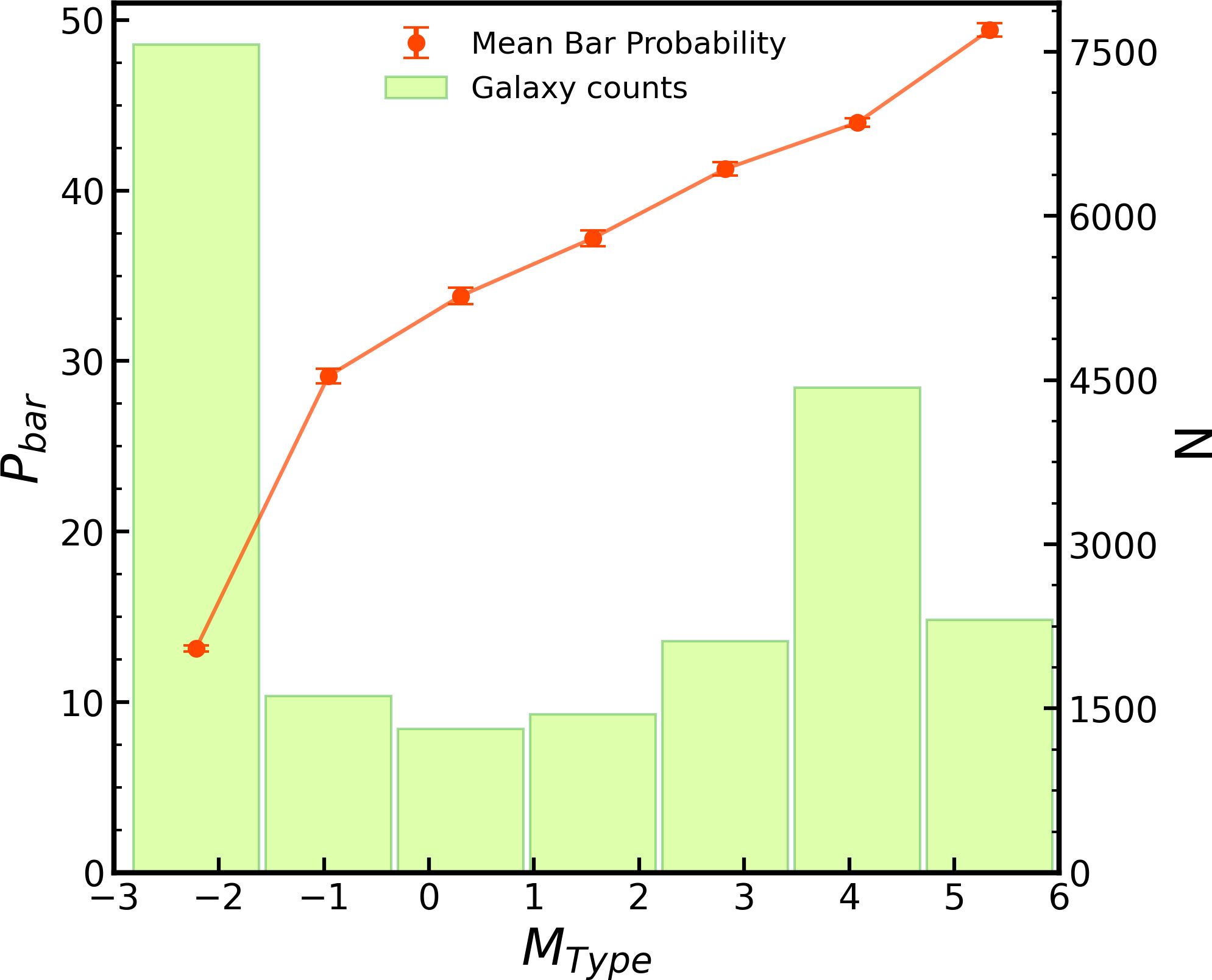}
    \caption{Probability of barred galaxies $P_{\mathrm{bar}}$ as a function of stellar mass $\log(M_{\ast}/M_{\odot})$ (left panel), colour $(g - r)$ (middle panel), and morphological type ($M_{\mathrm{type}}$; right panel).}
    \label{fig:fig1}
\end{figure*}

Environmental and physical properties were incorporated by matching the final
sample of 39,720 galaxies with the halo-based group catalogue of
\citet{Lim+2017}. This catalogue provides stellar masses, optical colours,
halo masses, and group membership information for SDSS galaxies.

In this work we adopt directly the stellar masses ($M_{\ast}$) and colours
$(g-r)$ provided in the galaxy catalogue associated with this dataset.
Stellar masses are estimated from $r$-band absolute magnitudes and
$(g-r)$ colours using the colour--mass-to-light ratio relation of
\citet{Bell+2003}. The photometric measurements are based on SDSS DR13
\citep{Albareti+2017}, including $K$- and evolutionary corrections to
$z=0.1$ following \citet{Poggianti+1997}, as implemented in
\citet{Lim+2017}. For galaxies with outlier colours, median values at
fixed luminosity were adopted.

Environmental parameters were also taken from the same halo-based group
catalogue, which was constructed using an improved group finder relative to
earlier implementations \citep{Yang+2005,Yang+2007}. In this framework,
halo masses are assigned through abundance matching using a proxy that
depends on both the stellar mass of the central galaxy and the stellar mass gap between the central and the $n$-th brightest satellite. Tests with mock catalogues show that this procedure reproduces true halo masses without systematic bias, with a typical uncertainty of $\sim$0.2 dex \citep{Lim+2017}.

For each group, the catalogue provides halo mass $M_{\mathrm{halo}}$,
group redshift, central and satellite identifiers, number of members,
and group-centric coordinates. Using the halo masses and group redshifts
provided in the catalogue, the corresponding virial radii were computed
following the definition adopted in \citet{Lim+2017}:

\begin{equation}
R_{180} = 1.33\,h^{-1}\,\mathrm{Mpc}
\left(\frac{M_{\mathrm{halo}}}{10^{14}\,h^{-1}\,M_{\odot}}\right)^{1/3}
(1 + z_{\mathrm{group}})^{-1}.
\end{equation}

The projected group-centric distance of each galaxy was then normalized as
$R/R_{180}$, providing a scale-independent measure of galaxy location within
haloes of different masses and redshifts. Central and satellite
classifications were directly adopted from \citet{Lim+2017}, where centrals
are defined as the most massive galaxies within each halo.

To explore environmental trends, galaxies were subdivided into three
categories: (i) isolated systems, (ii) satellites in low-richness groups with
fewer than $50$ members ($N < 50$), and (iii) satellites in high-richness
groups with $N \geq 50$ members. This classification enables a systematic
comparison of galaxy properties across a broad environmental range.

\subsection{Probabilistic Estimation of the Bar Likelihood}\label{subsec:probar}

Morphological classifications were taken from the catalogue of \citet{DominguezSanchez+2018} (DS18 hereafter), which provides automated classifications for SDSS DR7 galaxies using deep-learning models trained on visual morphological catalogues. The catalogue includes models based on both Galaxy Zoo 2 \citet{Willett+2013} and the visual classifications of \citet{Nair+2010a} (NA10 hereafter). In the present work, we adopt the T-type parameter and the lenticular probability, $P(\mathrm{S0})$, obtained from the models trained using the NA10 classifications. We also adopt the corresponding bar probability from the same NA10-based models. Thus, although Galaxy Zoo 2 forms part of the broader methodological basis of the DS18 catalogue, its classifications are not used directly in the present analysis. These outputs provide a homogeneous characterization of galaxy morphology and bar likelihood across the SDSS sample analysed here.

The T-type parameter spans approximately $-3$ to $6$ in the range covered by our sample and traces the sequence from early-type to late-type morphologies. Following the T-type convention adopted by DS18, non-positive T-type values correspond to the early-type regime, whereas progressively higher T-type values trace later morphological types. In our analysis, this parameter is used to characterize the broad morphological sequence of the galaxies.

To further characterize the lenticular component of the early-type
population, we use the additional lenticular probability, $P(\mathrm{S0})$,
provided by the same catalogue. DS18 developed this probabilistic
classification to improve the separation between elliptical and lenticular
galaxies within the early-type regime, and noted that it is meaningful for
galaxies with $T\leq0$. When morphology is taken into accountWhen morphology is taken into accountWhen morphology is taken into accountWe therefore identify galaxies with
$P(\mathrm{S0})>0.80$ as S0 systems for the purpose of the dedicated S0
analysis, while retaining S0 galaxies within the early-type population in
our main early/late classification. The sensitivity of this treatment
is examined in Appendix~\ref{app:s0_sensitivity}.

When morphology is taken into accountWhen morphology is taken into accountWhen morphology is taken into account
Bar classifications are also adopted from the same catalogue DS18. Specifically, we use the bar probability
obtained from the dedicated model trained on the visual bar identifications
of NA10. This provides a homogeneous probabilistic estimate
of bar likelihood across the full sample and is used consistently throughout
the present analysis.

Before applying the inclination cut, the parent sample contains 39,720 galaxies with valid morphological classifications. Of these, 14,013 galaxies have $T\leq0$ and are classified as early-type, while 25,707 have $T>0$ and are classified as late-type. Among the early-type population, 6,953 galaxies satisfy $P(\mathrm{S0})>0.80$ and are therefore identified as lenticular (S0) galaxies.

Following \citet{Lee+2012}, we restrict the analysis to galaxies with $b/a>0.6$. Here, $b/a$ is taken from the structural parameters of the \citet{Meert+2015} catalogue. This criterion is applied to reduce projection effects that hinder reliable bar identification in highly inclined systems and is independent of the morphological classification described above. Applying this inclination cut reduces the parent sample to 20,895 galaxies, comprising 9,533 early-type and 11,362 late-type galaxies. Among the early-type galaxies in this final sample, 3,825 satisfy $P(\mathrm{S0})>0.80$ and are therefore classified as S0 galaxies.

After applying this cut, the bar likelihood is computed using a fully 
probabilistic approach based on the individual bar probabilities, 
$P_{\mathrm{bar},i}$, provided by the classifier. By construction, 
$P_{\mathrm{bar},i}$ lies in the interval $[0,1]$ and represents the 
probability that galaxy $i$ hosts a stellar bar. Rather than adopting a 
binary classification, each galaxy contributes to the estimator according 
to its individual probability.

For a subset of $N$ galaxies, the bar likelihood, $P_{\mathrm{bar}}$, is 
defined as the average of the individual probabilities $P_{\mathrm{bar},i}$, 
expressed as a percentage. This quantity can therefore be interpreted as 
the expected barred fraction for the corresponding galaxy subset.

This formulation preserves the continuous probabilistic information delivered 
by the classifier and avoids the information loss inherent to 
threshold-based classifications, while remaining statistically equivalent 
to the expectation value of a binary bar indicator.

Statistical uncertainties are estimated using a non-parametric bootstrap 
with 1000 resamplings. For each subset, we recompute 
$P_{\mathrm{bar}}$ in each bootstrap realisation and derive the 
confidence intervals from the 16th and 84th percentiles of the resulting 
distribution, corresponding to a $1\sigma$ interval.

\begin{figure*}
    \centering
    \includegraphics[width=0.32\textwidth]{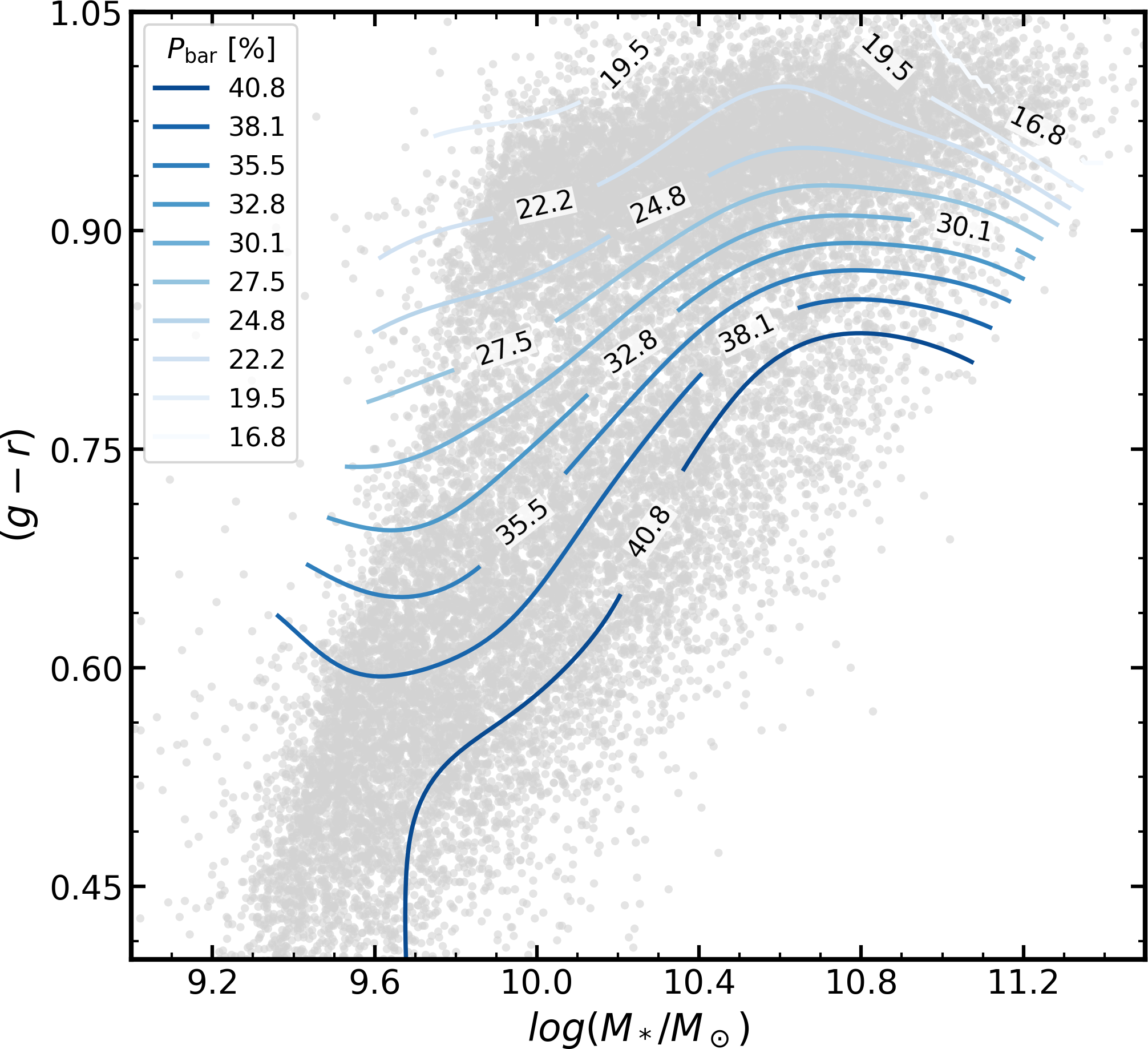}\hfill
    \includegraphics[width=0.32\textwidth]{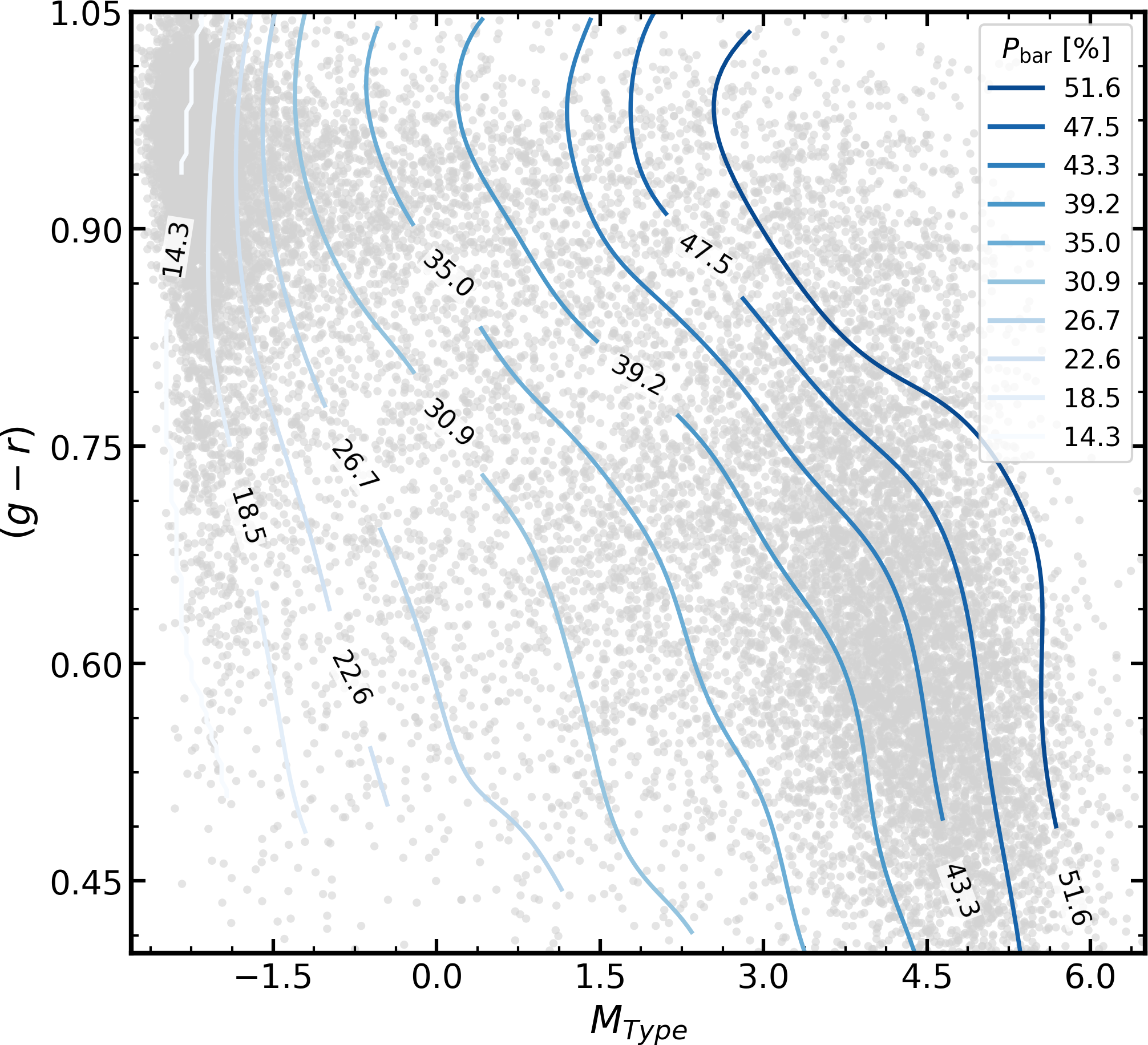}\hfill
    \includegraphics[width=0.32\textwidth]{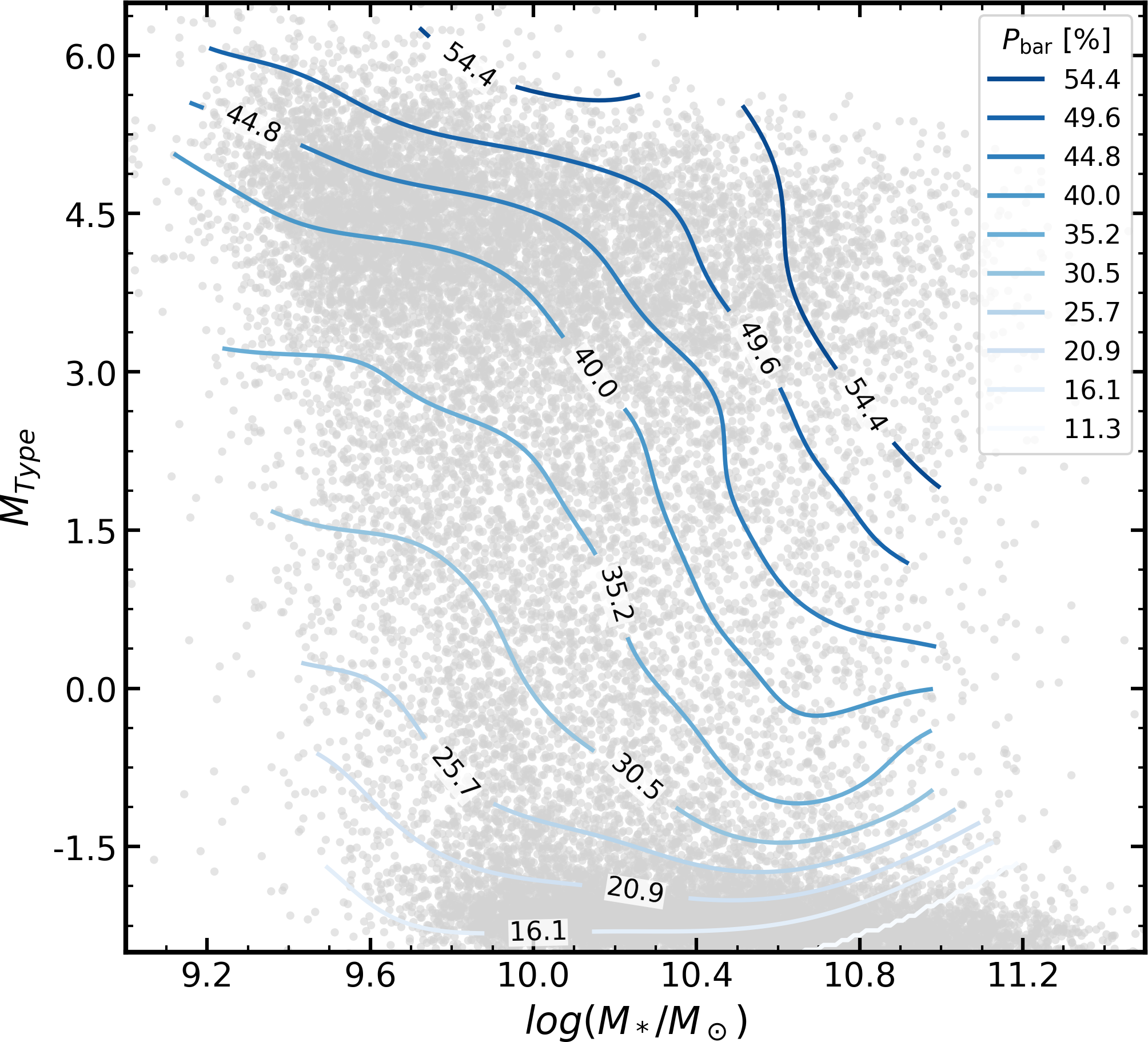}
    \caption{Probability of barred galaxies ($P_{\mathrm{bar}}$) shown as isocontours in the following parameter spaces:
    $(g - r)$ versus $\log(M_{\star}/M_{\odot})$ (left panel), $(g - r)$ versus $M_{\mathrm{type}}$ (middle panel), and $M_{\mathrm{type}}$ versus $\log(M_{\star}/M_{\odot})$ (right panel).
    The contours represent a smoothed estimate of the bar fraction and are shown only in regions of parameter space sufficiently populated to ensure robust statistics.}
    \label{fig:fig2}
\end{figure*}

\section{Results and Discussion}\label{sec:result}

In this section, we present the main results of our analysis on the incidence and properties of barred galaxies within the combined sample described in Section~\ref{sec:data}. The results are organized according to the principal parameters considered: stellar mass, morphology, colour and environment, as characterized by halo mass and group membership.

\subsection{Dependence of the Bar Likelihood on Intrinsic Galaxy Properties}\label{subsec:bar_intrinsic}

We begin our analysis by examining how the bar likelihood, $P_{\mathrm{bar}}$, varies with three key intrinsic galaxy parameters: total stellar mass ($M_{\ast}$), rest-frame colour $(g-r)$, and morphological type (T-type or $M_{\mathrm{type}}$).

In the left panel of Figure~\ref{fig:fig1}, we show $P_{\mathrm{bar}}$ as a function of total stellar mass. The data reveal a decreasing trend in bar likelihood with increasing stellar mass. Error bars represent the 1$\sigma$ confidence intervals derived from the bootstrap procedure described above. In all figures presented in this section, $P_{\mathrm{bar}}$ is computed using the probabilistic estimator defined in Section~\ref{subsec:probar}. The same procedure is consistently applied to all galaxy subsamples throughout the analysis.

This apparent decline of bar likelihood with stellar mass differs from several previous studies that reported either a flat or increasing bar fraction toward higher stellar masses \citep{Mendez-Abreu+2010, Lee+2012}. Another work has reported a more complex behaviour. In particular, \citet{Nair+2010b} found a bimodal distribution of the bar fraction as a function of stellar mass, with peaks at both low and high masses and a minimum at intermediate masses. Such a double-peaked trend is not recovered in our analysis. Instead, we find that the bar probability is primarily driven by morphological type, with late-type galaxies dominating the low-mass regime and early-type systems becoming more prevalent at higher masses, which naturally shapes the observed global trend. This global behaviour reflects changes in the morphological composition of the sample and differs from the trends observed at fixed morphology, where the bar likelihood increases with stellar mass.

A similar behaviour is observed in the central panel of Figure~\ref{fig:fig1}, where $P_{\mathrm{bar}}$ decreases toward redder $(g-r)$ colours.

However, the right panel of Figure~\ref{fig:fig1} provides key insight into the origin of these trends. We find a strong dependence of bar likelihood on morphological type, with $P_{\mathrm{bar}}$ increasing from early- to late-type galaxies. Early-type galaxies in our sample tend to be both more massive and redder, while simultaneously exhibiting a lower bar likelihood. Therefore, the apparent trends with stellar mass and colour are largely driven by the underlying correlation between these parameters and galaxy morphology.

This interpretation is consistent with studies using central concentration as a proxy for bulge prominence, which show a steady decline of bar fraction with increasing concentration \citep{Barazza+2008, Chacon+2024}. Both results point toward a common physical picture in which bars form and persist more efficiently in dynamically cold, rotationally supported discs, whereas a prominent bulge or high central concentration can stabilise the disc against bar instabilities \citep{Saha+2018}.

Overall, our results confirm that bar likelihood is governed by a combination of stellar mass, stellar population properties, and internal structure. However, since these parameters are not mutually independent, it is necessary to explore their joint effects in order to disentangle their relative contributions.

To investigate this interdependence, we examine the co-variation of the bar likelihood in two-dimensional parameter spaces, shown in Figure~\ref{fig:fig2}. For each parameter space, we calculate the mean $P_{\mathrm{bar}}$ on a regular two-dimensional grid and apply Gaussian smoothing to obtain a continuous representation of the bar likelihood. The resulting smoothed distributions are shown as contours in Figure~\ref{fig:fig2}. The two-dimensional contour maps presented throughout this work are intended primarily as descriptive visualizations of the joint variation of \(P_{\rm bar}\) across parameter space. They are used to identify patterns and trends that motivate the subsequent analysis, but are not interpreted by themselves as statistical tests of the significance of a given dependence. For the one-dimensional distributions, in contrast, the uncertainties shown in the figures are the \(1\sigma\) confidence intervals derived from the bootstrap procedure described above. The quantitative assessment of residual environmental dependence, after accounting for stellar mass and morphology, is presented separately in Section~\ref{subsec:residual_environment}.

Figure~\ref{fig:fig2} presents isocontours of the smoothed bar likelihood in three parameter planes: $(g - r)$ versus $\log(M_{\ast}/M_{\odot})$ (left panel), $M_{\mathrm{type}}$ versus $\log(M_{\ast}/M_{\odot})$ (middle panel), and $(g - r)$ versus $M_{\mathrm{type}}$ (right panel). In this context, $P_{\mathrm{bar}}$ represents the expected barred fraction at a given location in parameter space.

In the $(g - r)$--$\log(M_{\ast}/M_{\odot})$ plane, the bar likelihood increases toward more massive and bluer galaxies. The isocontours indicate that the dependence on stellar mass and colour is not independent, reflecting the well-known covariance between stellar mass and stellar population properties in late-type galaxies. Similar coupled trends between stellar mass, colour, and bar fraction have been reported in recent studies of nearby galaxy samples \citep{Mukundan+2025}.

In contrast, in the $(g - r)$--$M_{\mathrm{type}}$ plane, the dependence on colour becomes significantly weaker once morphological type is fixed. This demonstrates that the apparent colour dependence observed in Figure~\ref{fig:fig1} is largely driven by the underlying correlation between colour and morphology. Moreover, at fixed morphological type, the trend with colour partially reverses, with $P_{\mathrm{bar}}$ showing a mild increase toward redder systems. This behaviour indicates that colour alone is not the primary driver of bar likelihood, but rather reflects the structural properties encoded in galaxy morphology.
Finally, in the $M_{\mathrm{type}}$--$\log(M_{\ast}/M_{\odot})$ plane, the bar 
likelihood for late-type galaxies remains approximately constant from 
$M_{\mathrm{type}} \sim 0$ toward later types for stellar masses 
$M_{\ast} \lesssim 10^{10.5}\,M_{\odot}$, increasing only at higher masses. 
In contrast, early-type galaxies exhibit systematically lower values of 
$P_{\mathrm{bar}}$ across the entire stellar mass range, with only a weak 
dependence on mass. 
These results indicate that the probability of hosting a bar is not determined 
by a single parameter, but instead arises from the combined effect of morphology 
and stellar mass, with morphology setting the primary baseline and stellar mass 
modulating the bar likelihood within each morphological class.
These results indicate that bar likelihood is jointly regulated by 
morphological type and stellar mass. While the strongest gradients in 
$P_{\mathrm{bar}}$ reflect a combined dependence on both parameters, 
the behaviour within fixed morphological classes reveals additional trends. 
Once morphology is fixed, stellar mass plays a secondary role, whereas 
early-type systems maintain a low probability of hosting a bar largely 
independent of stellar mass. 
This behaviour reinforces the interpretation that internal structure — 
particularly bulge prominence and disc stability — acts as the primary 
regulator of bar formation. The mass-dependent enhancement observed in 
massive late-type galaxies may reflect the increasing importance of disc 
self-gravity in sustaining long-lived bar instabilities.

\begin{table}
    \centering
    \caption{Mean bar probability $P_{\mathrm{bar}}$ for the different galaxy subsets analysed in this work.}
    \label{tab:pbar_results}
    \begin{tabular}{lcc}
        \hline
        Subset & $N$ & $P_{\mathrm{bar}}$ (\%) \\
        \hline
        All field galaxies     & 10869 & $33.20 \pm 0.20$ \\
        All central galaxies   & 3238  & $27.06 \pm 0.46$ \\
        All satellite galaxies & 6788  & $29.21 \pm 0.25$ \\
        \hline
        Field late-type        & 6845  & $42.15 \pm 0.22$ \\
        Field early-type       & 4024  & $17.98 \pm 0.27$ \\
        \hline
        Central late-type      & 1372  & $49.38 \pm 0.59$ \\
        Central early-type     & 1866  & $10.65 \pm 0.39$ \\
        \hline
        Satellite late-type    & 3145  & $42.19 \pm 0.33$ \\
        Satellite early-type   & 3643  & $18.00 \pm 0.29$ \\
        \hline
    \end{tabular}
\end{table}

\subsection{Environmental and Global Trends in the Bar Probability}

We now explore whether the bar likelihood exhibits systematic variations with halo-scale environment once galaxy morphology is taken into account. Using the probabilistic estimator defined in Section~\ref{subsec:probar}, we compare $P_{\mathrm{bar}}$ across different environmental regimes.

Table~\ref{tab:pbar_results} summarises the mean bar probability for the full sample and for galaxies classified as field, centrals, and satellites, further divided into late- and early-type systems. When all morphological types are combined, field galaxies exhibit a mean bar likelihood of $P_{\mathrm{bar}} = 33.20 \pm 0.20\%$, while central and satellite galaxies show comparable values of $27.06 \pm 0.46\%$ and $29.21 \pm 0.25\%$, respectively.
These values are consistent with previous measurements of bar fractions in local galaxies \citep[e.g.,][]{Lin+2014,Chacon+2025}, which report typical fractions of $\sim 30$--40\% for $\log_{10}(M_*/M_\odot) > 10$. In particular, \citet{Chacon+2025} derived similar values using a representative sample of nearby field galaxies, providing a direct local benchmark for our results.
For completeness, we note that our measurements are also in agreement with the global bar fraction reported by \citet{Euclid+2026}. However, since that study does not distinguish between centrals, satellites, and field galaxies, the comparison is restricted to the overall bar incidence.
When galaxies are separated by morphological type, a clear trend emerges: late-type galaxies show the highest bar probabilities, reaching $P_{\mathrm{bar}} \sim 42\%$ for field and satellite systems and up to $49\%$ for central late-types, while early-type galaxies exhibit significantly lower values ($\sim 11$--18\%) across all environments. This indicates that morphology is the primary driver of bar occurrence, with only a secondary dependence on environment.

The results reveal a strong dependence of $P_{\mathrm{bar}}$ on galaxy morphology across all environments. Late-type galaxies consistently exhibit significantly higher bar probabilities than early-type galaxies, regardless of whether they reside in the field, are central galaxies, or satellites. In contrast, early-type galaxies show uniformly low bar probabilities, indicating that the presence of a dominant bulge or dynamically hot stellar component suppresses bar formation.

At fixed morphology, the dependence of $P_{\mathrm{bar}}$ on environment is comparatively weak. Late-type central galaxies display the highest bar probability ($P_{\mathrm{bar}} \simeq 49\%$), while late-type field and satellite galaxies show very similar values ($P_{\mathrm{bar}} \simeq 42\%$). A similar pattern is observed for early-type galaxies, for which the bar probability remains low and nearly independent of environment. These results indicate that environment plays a secondary role in the observed bar probability. Most of the environmental dependence appears to arise through its correlation with galaxy morphology, although weaker direct environmental effects cannot be excluded within the sensitivity of the present analysis.

Our findings are consistent with previous observational studies showing that bar incidence is primarily governed by internal galaxy properties, with only a weak or indirect dependence on environment \citep{Skibba+2012, Lin+2014, Erwin+2018}. The enhanced bar probability observed in late-type central galaxies may reflect the combined effects of higher stellar mass and disc self-gravity, favouring the growth and long-term survival of bar instabilities. The agreement with previous studies supports the robustness of our bar identification, as our sample reproduces well-established trends. We therefore explore in the following sections the combined role of intrinsic galaxy properties and environment in shaping bar formation and evolution.

\begin{figure}
    \centering
    \includegraphics[width=\columnwidth]{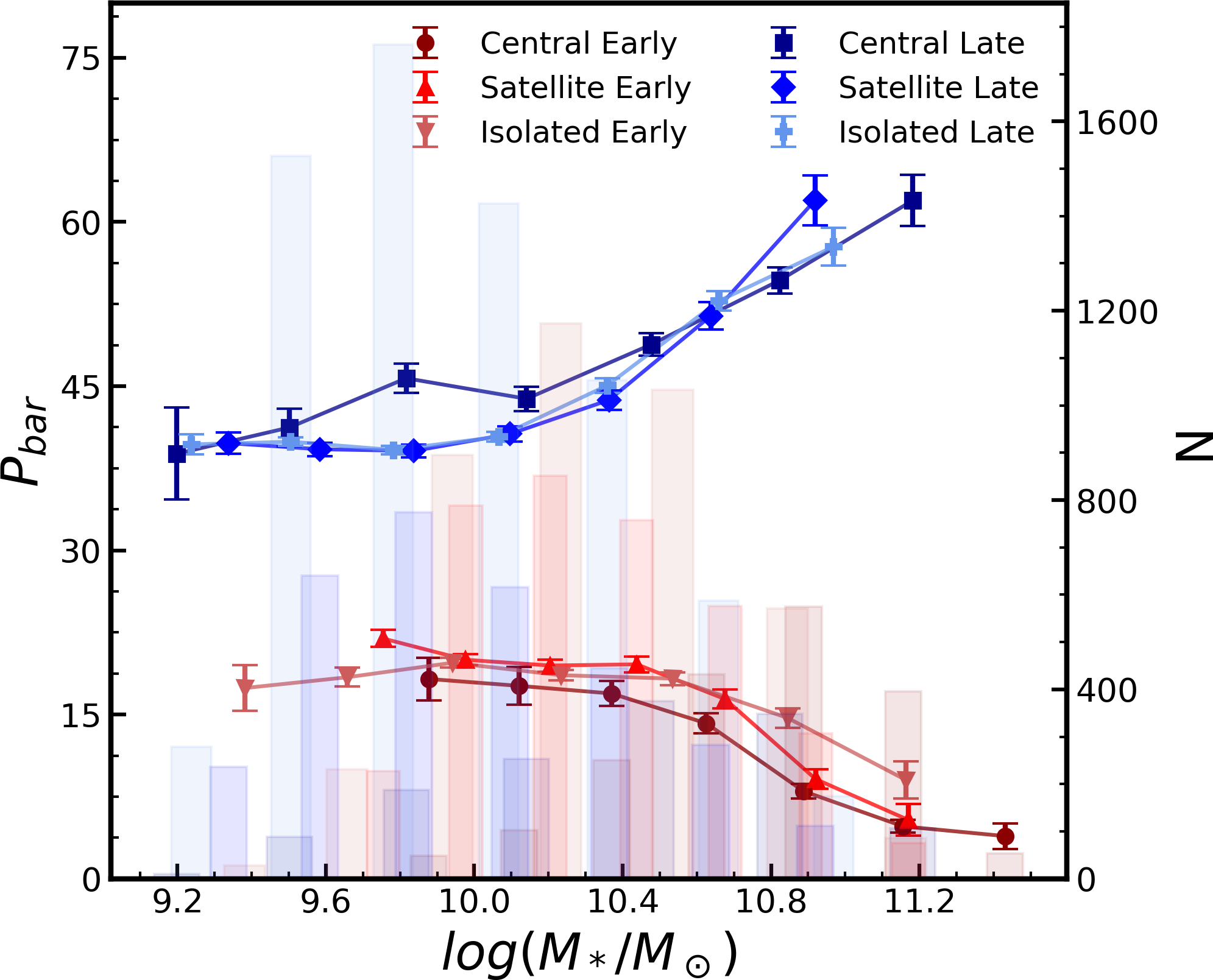}
    \caption{Mean bar likelihood, $P_{\mathrm{bar}}$, as a function of stellar mass $\log(M_{\ast}/M_{\odot})$ for early- and late-type galaxies in different environmental regimes: central, satellite, and isolated systems.}
    \label{fig:fig3}
\end{figure}

\subsection{Stellar mass effects at fixed morphology in different environments}

\begin{figure*}
    \centering
    \includegraphics[width=0.32\textwidth]{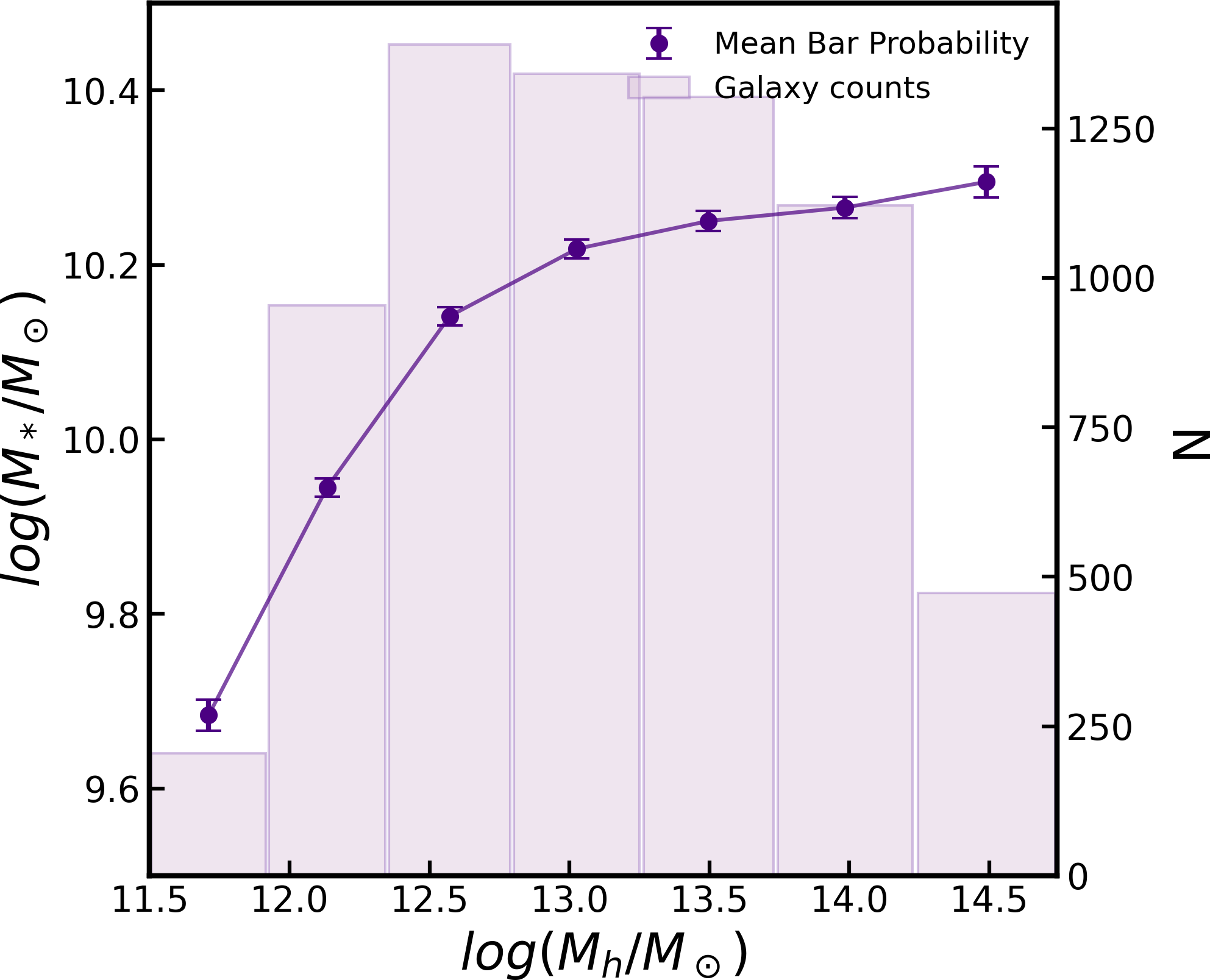}\hfill
    \includegraphics[width=0.32\textwidth]{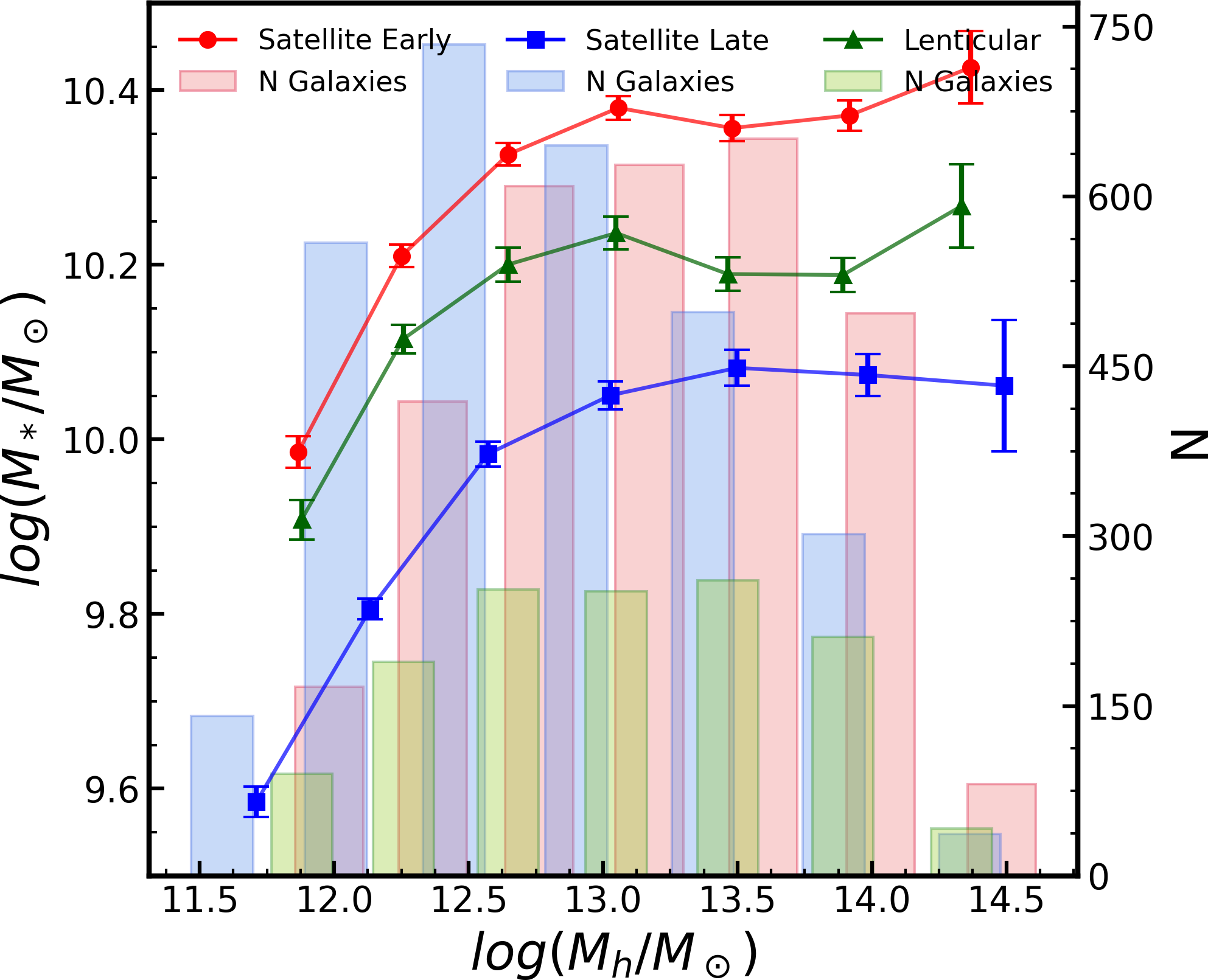}\hfill
    \includegraphics[width=0.32\textwidth]{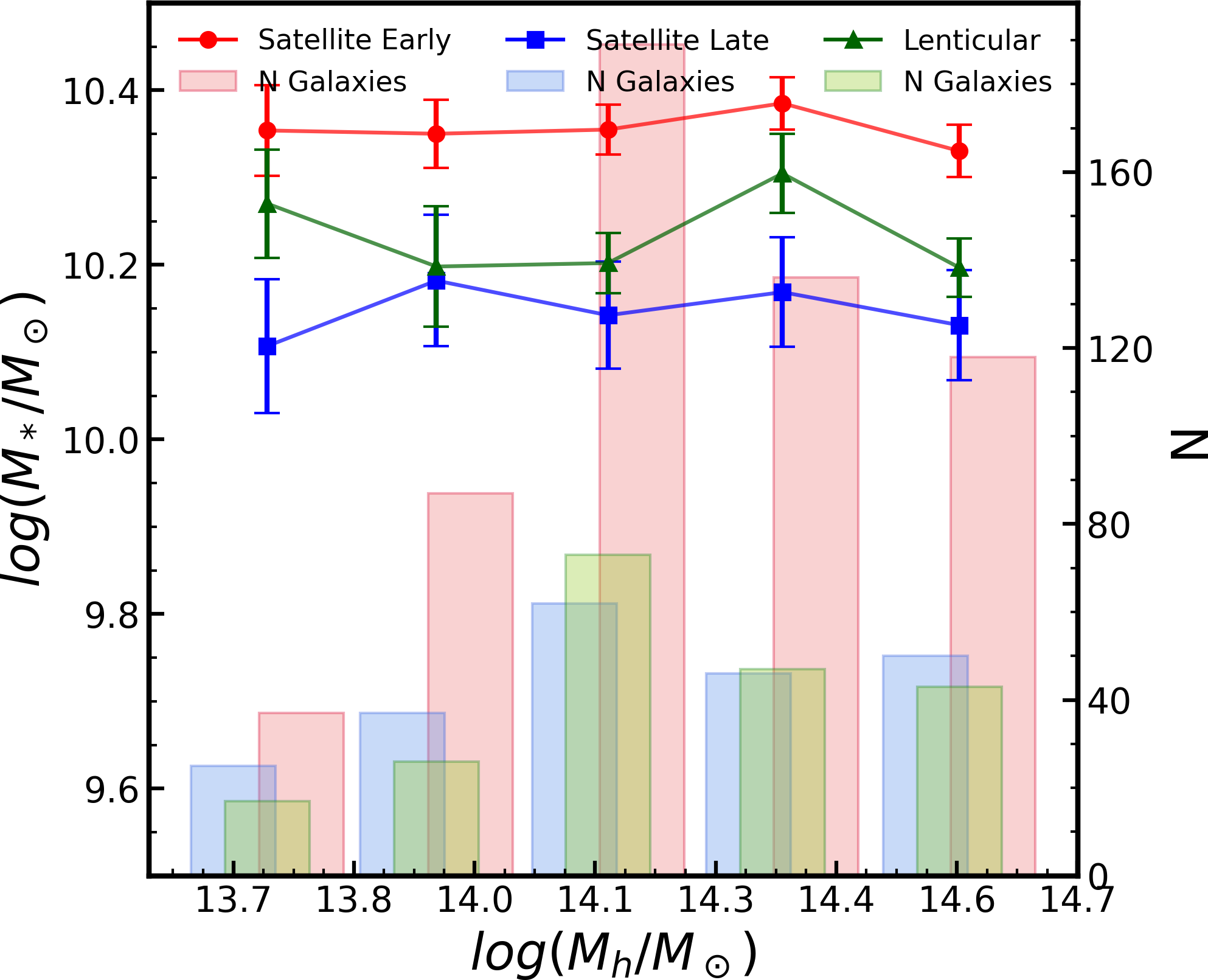}
    \caption{Stellar mass $\log(M_{\ast}/M_{\odot})$ as a function of halo mass $\log(M_{h}/M_{\odot})$ for satellite galaxies.
    Left panel: full satellite sample.
    Middle panel: satellites ($N < 50$).
    Right panel: satellites ($N \geq 50$).
    Middle and right panels show early-type, late-type, and lenticular (S0, $P(\mathrm{S0}) > 0.80$) satellites separately.}
    \label{fig:fig4}
\end{figure*}

Table~\ref{tab:pbar_results} shows that late-type central galaxies exhibit a higher mean bar likelihood than their isolated counterparts. Since central late-type galaxies tend to be, on average, more massive than isolated late-type systems, it is necessary to control for stellar mass in order to determine whether this difference is genuinely environmental or simply reflects underlying mass variations. For this reason, we examine the mass dependence of $P_{\mathrm{bar}}$ separately for each environmental class.

To explore the dependence of the bar likelihood on environment, we analyse the behaviour of $P_{\mathrm{bar}}$ as a function of stellar mass, $\log(M_{\ast}/M_{\odot})$, enabling a controlled comparison at fixed mass. The sample is divided into central, satellite, and isolated galaxies, and each environmental category is further split into early- and late-type systems. This approach allows us to isolate residual environmental trends in $P_{\mathrm{bar}}$ while accounting for differences associated with galaxy morphology.

Figure~\ref{fig:fig3} presents the resulting trends. For late-type galaxies, the bar likelihood increases with stellar mass in all environments — centrals, satellites, and isolated systems — following a similar behaviour across the three categories. In contrast, early-type galaxies display systematically low bar likelihood values over the entire mass range, with a weak or mildly decreasing dependence on stellar mass, independently of environment. This behaviour does not represent a general discrepancy with previous studies 
that reported a positive correlation between bar fraction and stellar mass 
\citep{Skibba+2012, Gavazzi+2015}. Instead, our results show that this trend 
is recovered for late-type galaxies, where $P_{\mathrm{bar}}$ increases with 
stellar mass. In contrast, early-type systems exhibit systematically lower 
bar probabilities with only a weak dependence on stellar mass, which explains 
the apparent differences when considering the full galaxy population.
This highlights the importance of separating galaxy populations by morphology 
when interpreting global trends.

These results suggest that environmental effects are not the primary driver of bar formation. At fixed stellar mass and morphological type, the probability of hosting a bar appears largely insensitive to whether a galaxy is central, satellite, or isolated. Instead, morphology remains the dominant parameter regulating bar likelihood. It is important to note, however, that morphology itself is unlikely to be the fundamental physical driver. Rather, morphological type probably acts as an observational proxy for the underlying dynamical state of the galaxy, encapsulating structural and kinematic properties such as bulge prominence, disc stability, stellar angular momentum, and gas content, all of which are expected to influence the susceptibility of galactic discs to bar instabilities. \citep{Rosas-Guevara22, Lu25, Chim25}

\begin{figure}
    \centering
    \includegraphics[width=0.39\textwidth]{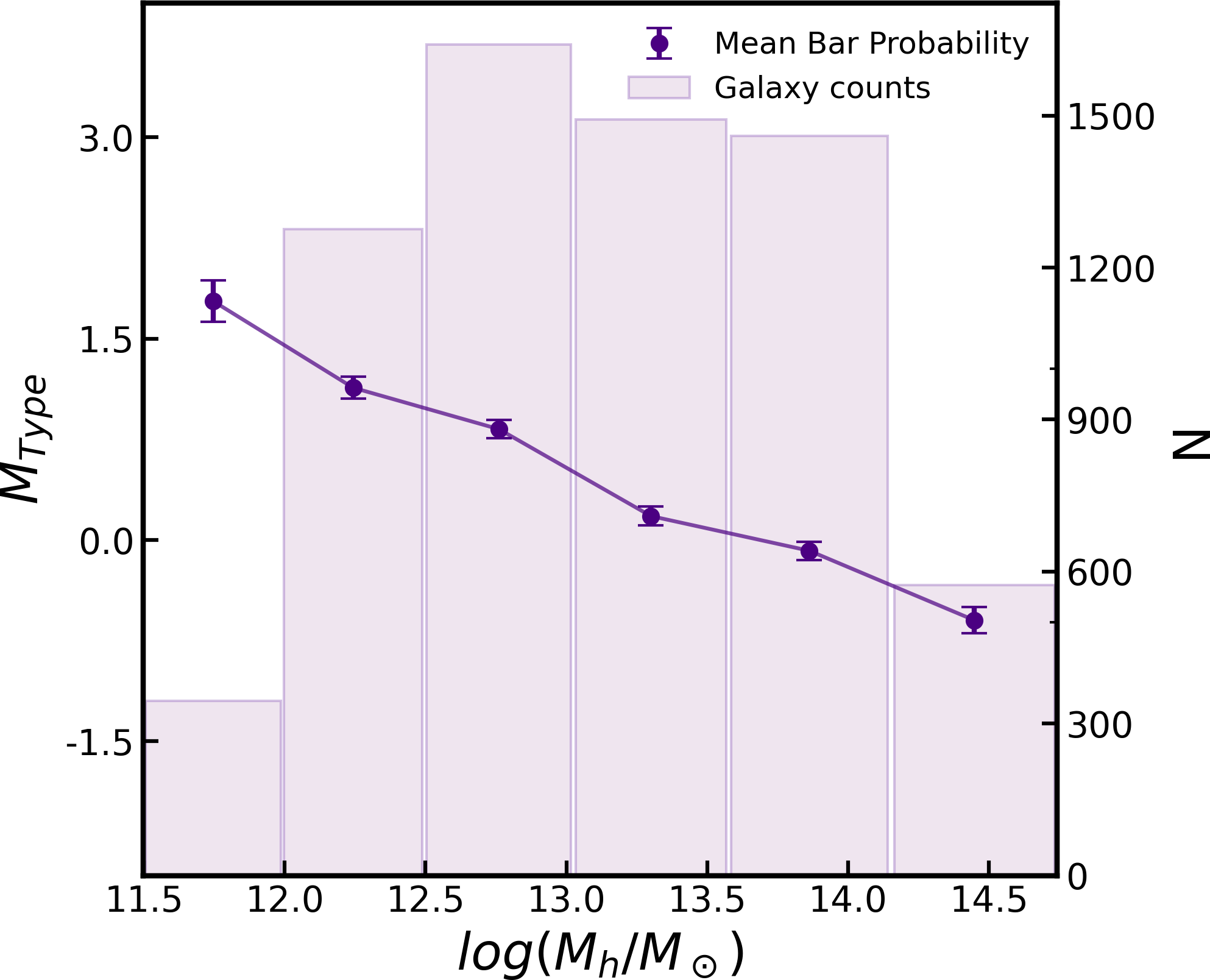}
    
    \includegraphics[width=0.38\textwidth]{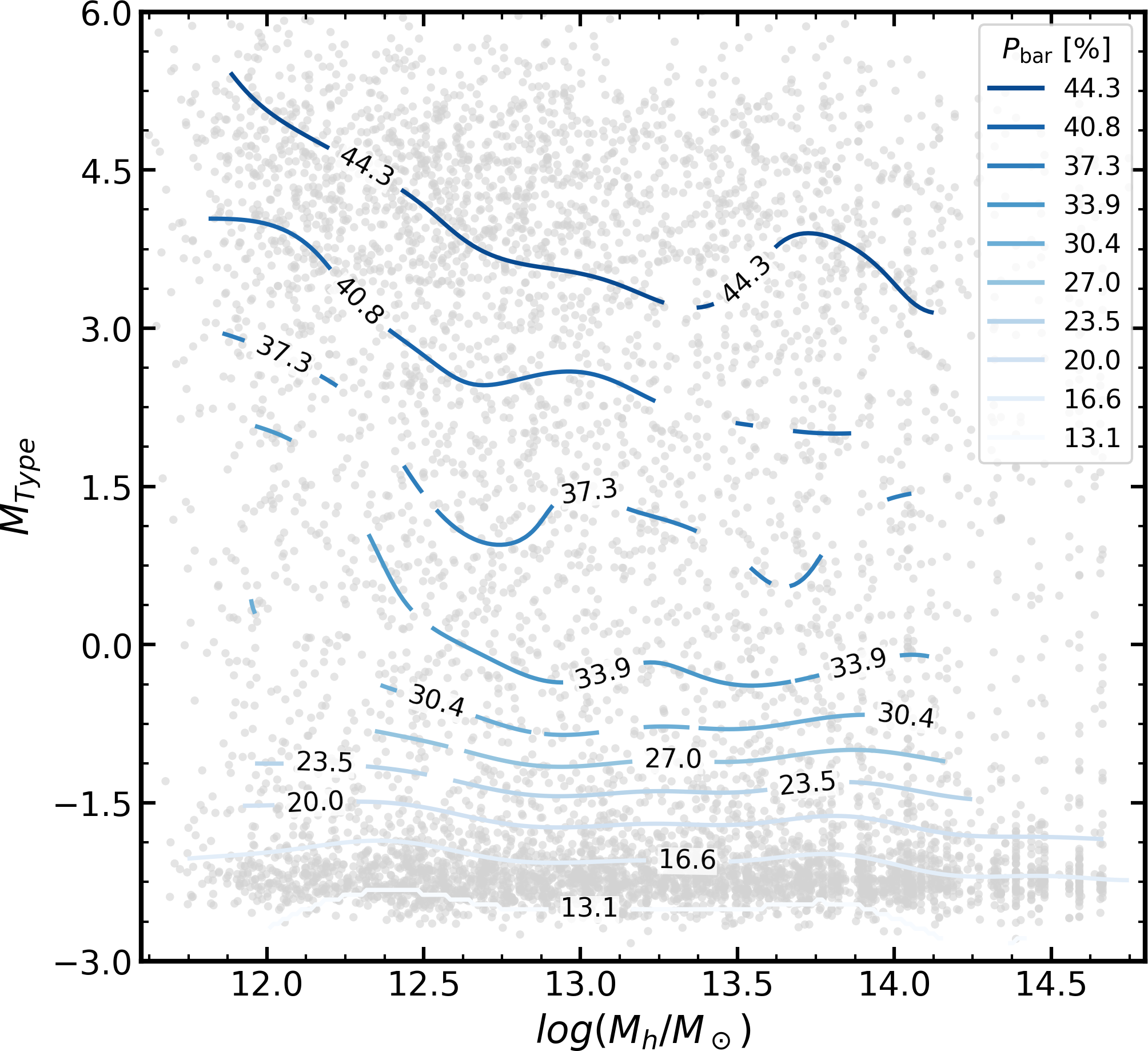}
    
    \caption{Morphological type ($M_{\mathrm{type}}$) as a function of halo mass $\log(M_{\mathrm{h}}/M_{\odot})$ for satellite galaxies.
    Top panel: the full satellite sample.
    Bottom panel: all satellite galaxies, where the blue line denotes the isocontour of the probability of being a barred galaxy.}
    \label{fig:fig5}
\end{figure}

\subsection{Halo mass and the morphology-driven origin of environmental trends}
\label{subsec:shm_satellites}

\begin{figure*}
    \centering
    \includegraphics[width=0.48\textwidth]{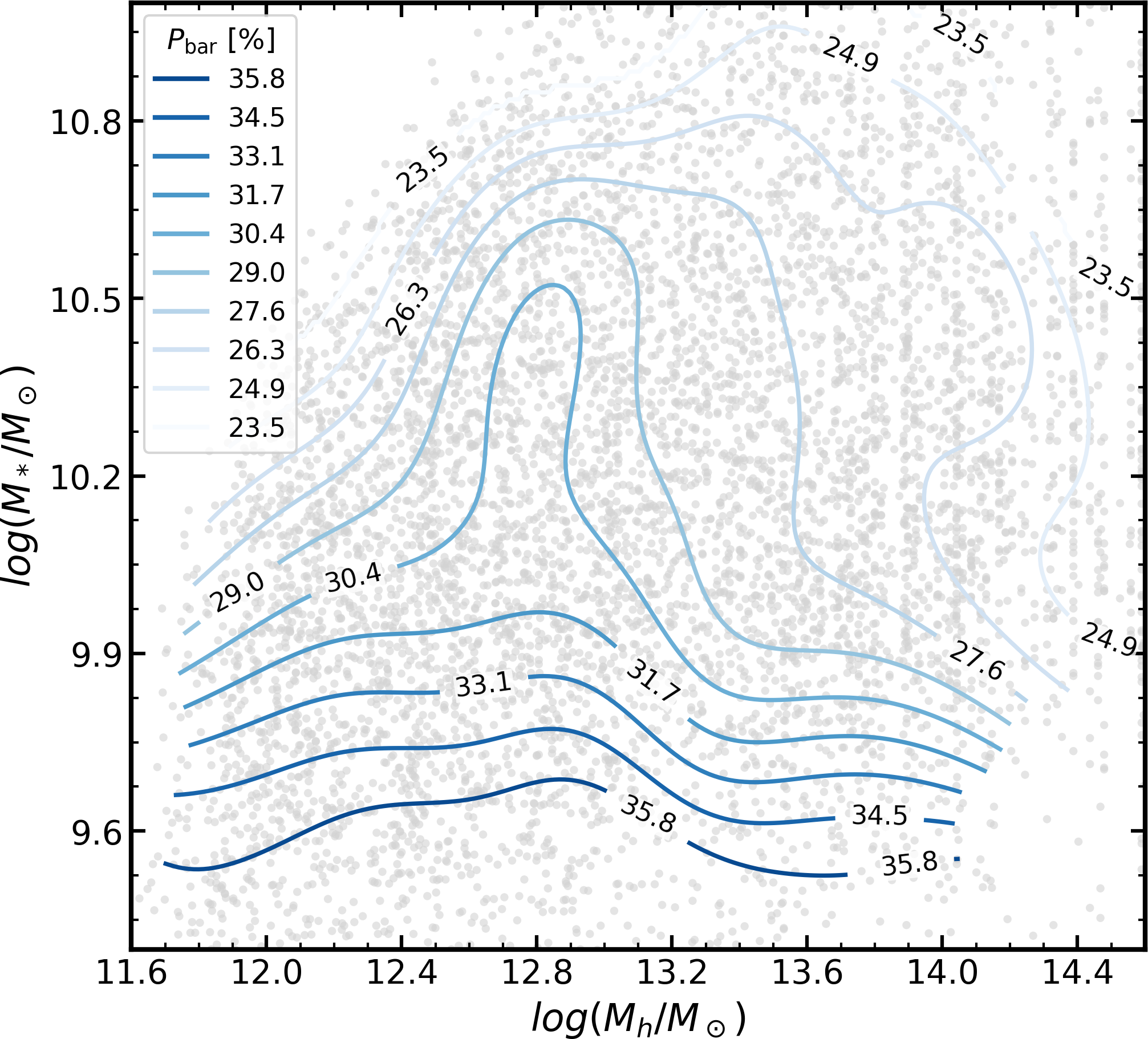}\hfill
    \includegraphics[width=0.48\textwidth]{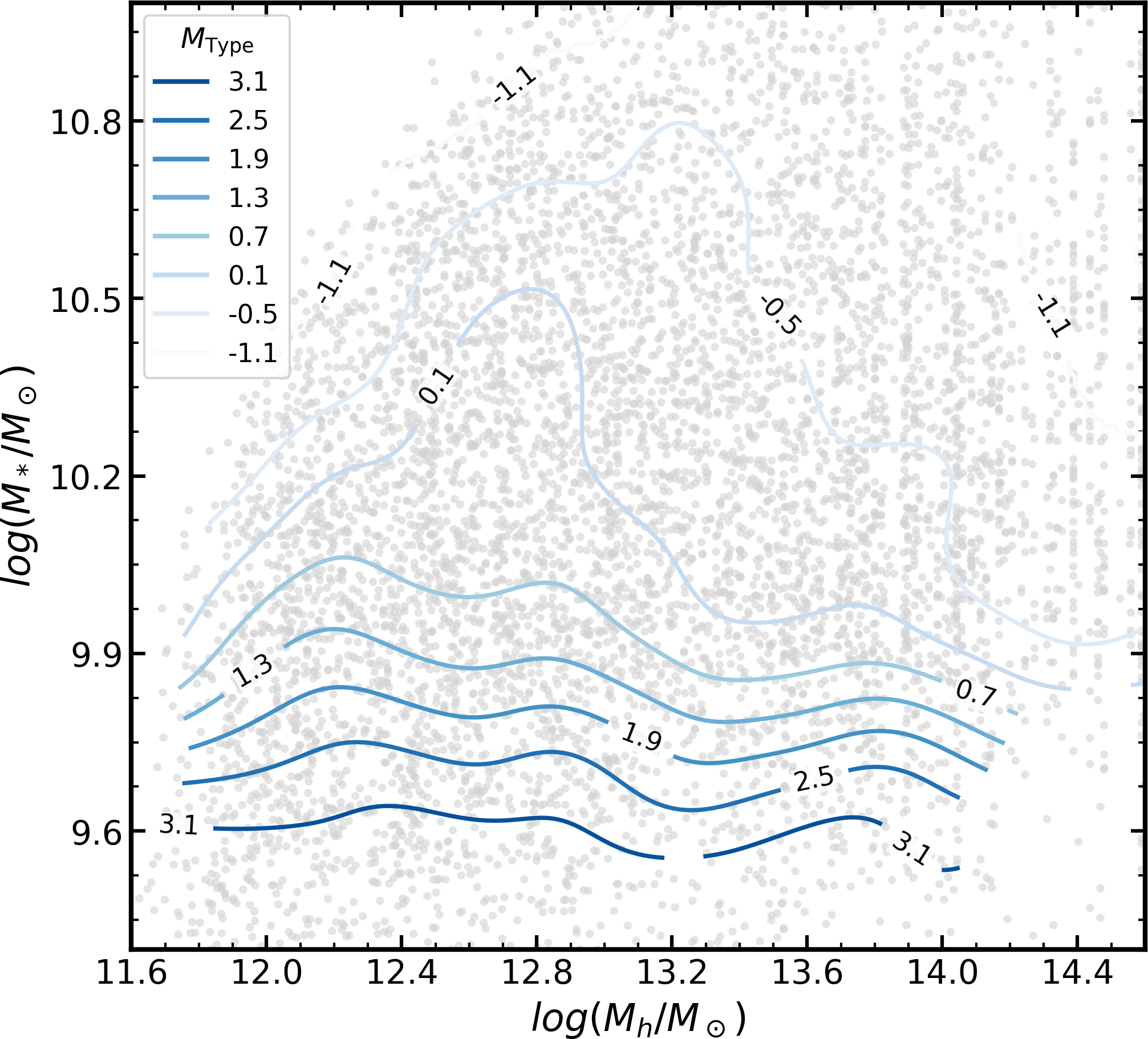}
    \caption{Isocontours in the $\log(M_{\star}/M_{\odot})$ versus $\log(M_{\mathrm{h}}/M_{\odot})$ parameter space showing:
    the probability of barred galaxies ($P_{\mathrm{bar}}$) in the left panel,
    and the morphological type ($M_{\mathrm{type}}$) in the rigth panel.}
    \label{fig:fig6}
\end{figure*}

In the previous subsection, we showed that, at fixed stellar mass and morphological type, the bar likelihood is largely independent of environment. This suggests that there is no strong evidence that the halo-scale environment directly regulates bar formation once morphology is controlled for. However, at these scales, environment is primarily characterised by halo mass, which may indirectly influence bar likelihood by modulating the stellar mass distribution of galaxies residing in haloes of different mass.

To explore this possibility, we examine the relation between stellar mass and halo mass for satellite galaxies, which allows us to assess whether variations in bar incidence are linked to differences in halo mass at fixed stellar mass. Figure~\ref{fig:fig4} shows the distribution of satellites in the $\log(M_{\ast}/M_{\odot})$–$\log(M_{h}/M_{\odot})$ space, separated according to morphological class: early-type, late-type, and the lenticular subsample identified through their high S0 probabilities.

The left panel presents the full satellite population, while the middle and right panels separate satellites according to the richness of their host haloes: systems with $N < 50$ members (minor groups) and those with $N \geq 50$ members (major groups). This subdivision allows us to explore how the joint distribution of stellar mass, halo mass, and morphology varies with increasing halo mass.

A clear morphological segregation is observed in the stellar mass–halo mass relation. Satellite galaxies in more massive haloes tend to exhibit higher stellar masses on average, with this trend being more pronounced in richer environments. At fixed halo mass, early-type galaxies are typically more massive than lenticular systems, which in turn are more massive than late-type satellites.
These trends are consistent with previous studies showing that galaxy stellar mass correlates with both morphology and environment \citep{Pfeffer+2023}. We emphasise, however, that this figure reflects average mass trends rather than direct evidence of morphological transformation. Instead, it suggests that the observed differences in bar incidence may be partly linked to systematic variations in stellar mass across morphological types and environments.

Figure~\ref{fig:fig5} further illustrates this segregation by showing the distribution of 
morphological type, $M_{\rm type}$, as a function of halo mass for satellite galaxies. 
The top panel displays the overall morphology--halo mass relation for satellite galaxies. 
We observe that $M_{\rm type}$ decreases systematically with increasing halo mass, 
indicating a shift toward earlier morphological types in more massive haloes. 
This trend reflects the well-established morphology--environment relation, 
where dense environments preferentially host earlier-type systems 
\citep{Dressler+1980, Carollo+2013}.

The bottom panel overlays the isocontours of the mean bar likelihood, 
$P_{\mathrm{bar}}$, in the $M_{\rm type}$--$\log(M_h/M_{\odot})$ plane. 
The contours are predominantly horizontal, indicating that the bar probability depends 
primarily on morphological type rather than halo mass. At fixed morphology, the variation 
of $P_{\mathrm{bar}}$ with halo mass is weak, with only a mild modulation 
toward intermediate halo masses. 

This behaviour reinforces our previous result that the likelihood of hosting a bar is primarily associated with the internal structural properties of galaxies. Once morphology is taken into account, halo mass contributes only weakly to the observed trends. Nevertheless, our analysis cannot exclude the presence of more subtle environmental effects that may lie below the statistical sensitivity of the present data. This interpretation is consistent with recent observational and theoretical studies showing that bar formation is mainly governed by the internal dynamical conditions of galactic discs, while environmental processes act primarily as secondary triggers or influence the subsequent evolution of pre-existing bars \citep{Zana18, Aquino-Ortiz+2026}.

This interpretation is reinforced by the isocontour maps shown in Fig.~\ref{fig:fig6}, 
where the probability of hosting a bar, \(P_{\mathrm{bar}}\), and the morphological type 
are mapped simultaneously in the \((M_{\ast}, M_{h})\) plane. The regions of enhanced bar 
probability closely follow those occupied by disc-dominated and lenticular satellites, 
indicating that the dependence of bar incidence on halo mass is largely mediated through 
its impact on galaxy morphology and modulated by the stellar mass of the galaxies.

These results are in good agreement with recent observational studies showing that the 
halo-scale environment primarily influences the structural properties of galaxies rather 
than directly triggering bar formation. In particular, \citet{Aguerri+2023} demonstrated 
that, while bars in dense environments such as galaxy clusters tend to be shorter and 
weaker—especially in low-luminosity systems—their occurrence is mainly governed by internal 
galaxy properties. In this context, our findings suggest that the observed dependence of
bar probability on host halo mass is largely associated with its connection
to stellar mass and morphological composition, rather than providing evidence
for a direct suppression or enhancement of bar formation, consistent with
theoretical studies showing that bar evolution is strongly coupled to internal
angular-momentum exchange between the stellar disc and the surrounding
spheroidal components \citep{Athanassoula+2003}.

Similar trends have been reported in previous studies exploring the connection between 
satellite galaxy properties and their host haloes. The preferential location of early--type 
and lenticular satellites in massive haloes is consistent with the well-established 
morphology--density and morphology--halo mass relations 
\citep{Peng+2012}. 
In this framework, the enhanced bar probability observed at intermediate halo masses can be 
interpreted as a transitional regime, where environmental mechanisms have modified the 
global structure of galaxies without completely erasing their disc components. 
This scenario naturally explains why bar formation remains efficient in a subset of 
satellite galaxies, even in relatively dense environments, provided that dynamically cold 
discs are still present \citep{Tawfeek+2022}.

Interestingly, a local enhancement in $P_{\mathrm{bar}}$ around
$\log(M_{\mathrm{h}}/M_{\odot}) \sim 12.8$ is visible in the bar-likelihood
map of Fig.~\ref{fig:fig6} and coincides with the corresponding morphology
distribution in the paired panel. This feature may reflect the underlying
distribution of satellite galaxies across stellar mass and morphology at
intermediate halo masses. However, the residual analysis presented in
Section~\ref{subsec:residual_environment} does not reveal a statistically
significant systematic environmental dependence after accounting for stellar
mass and morphology. We therefore interpret this local enhancement as part
of the population structure rather than as evidence for a distinct halo-mass
scale governing bar formation.

Fig.~\ref{fig:fig6} provides one of the clearest pieces of evidence supporting the interpretation developed in this work. The close correspondence between the isocontours of $P_{\mathrm{bar}}$ in the left panel and the distribution of morphological type, $M_{\mathrm{Type}}$, in the right panel shows that the regions of highest bar likelihood are predominantly occupied by galaxies with later morphological types. Conversely, regions characterized by lower $P_{\mathrm{bar}}$ correspond to progressively earlier morphological types. This correspondence persists across the stellar-mass--halo-mass plane and indicates that the variation of bar likelihood with halo mass is closely connected to the changing morphological composition of the galaxy population.

In particular, the figure illustrates that halo mass and stellar mass do not populate the parameter space independently of morphology. At a given range of halo mass, galaxies occupy different regions of the plane according to their morphological type, and these regions are associated with systematically different values of $P_{\mathrm{bar}}$. Thus, the apparent environmental dependence of bar likelihood can arise naturally from the different morphological populations that occupy different locations in the stellar-mass--halo-mass plane. Figure~\ref{fig:fig6} therefore provides an important visual demonstration of why the environmental trends observed in $P_{\mathrm{bar}}$ should not be interpreted, by themselves, as evidence for a direct causal effect of halo mass on bar formation.

The quantitative residual analysis presented in Section~\ref{subsec:residual_environment} provides a complementary test of this interpretation by explicitly accounting for stellar mass and morphology. Together, the correspondence shown in Fig.~\ref{fig:fig6} and the residual analysis indicate that the observed halo-mass dependence of $P_{\mathrm{bar}}$ is largely associated with the underlying stellar-mass and morphological composition of the galaxy population, while no statistically significant residual dependence on halo mass is detected within the sensitivity of the present sample.

Having characterised how stellar mass and morphology vary with halo mass for satellite galaxies, we now examine whether the bar likelihood retains any residual dependence on halo mass after accounting for these galaxy properties. In particular, we compare satellite and central galaxies to determine how the observed halo-mass trends vary across the two populations and whether they persist once morphology and stellar mass are taken into account.

\begin{figure*}
    \centering
    \includegraphics[width=0.32\textwidth]{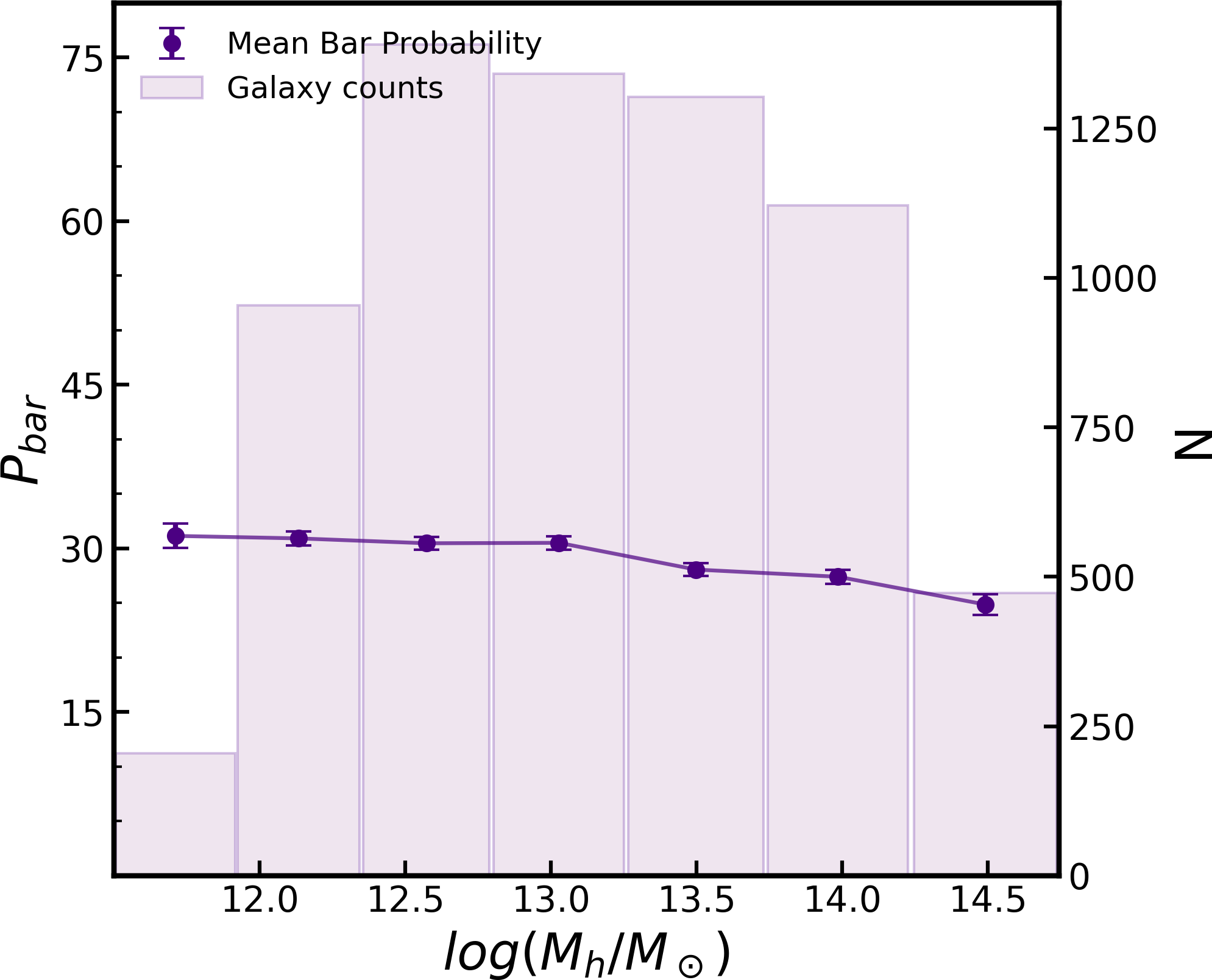}\hfill
    \includegraphics[width=0.32\textwidth]{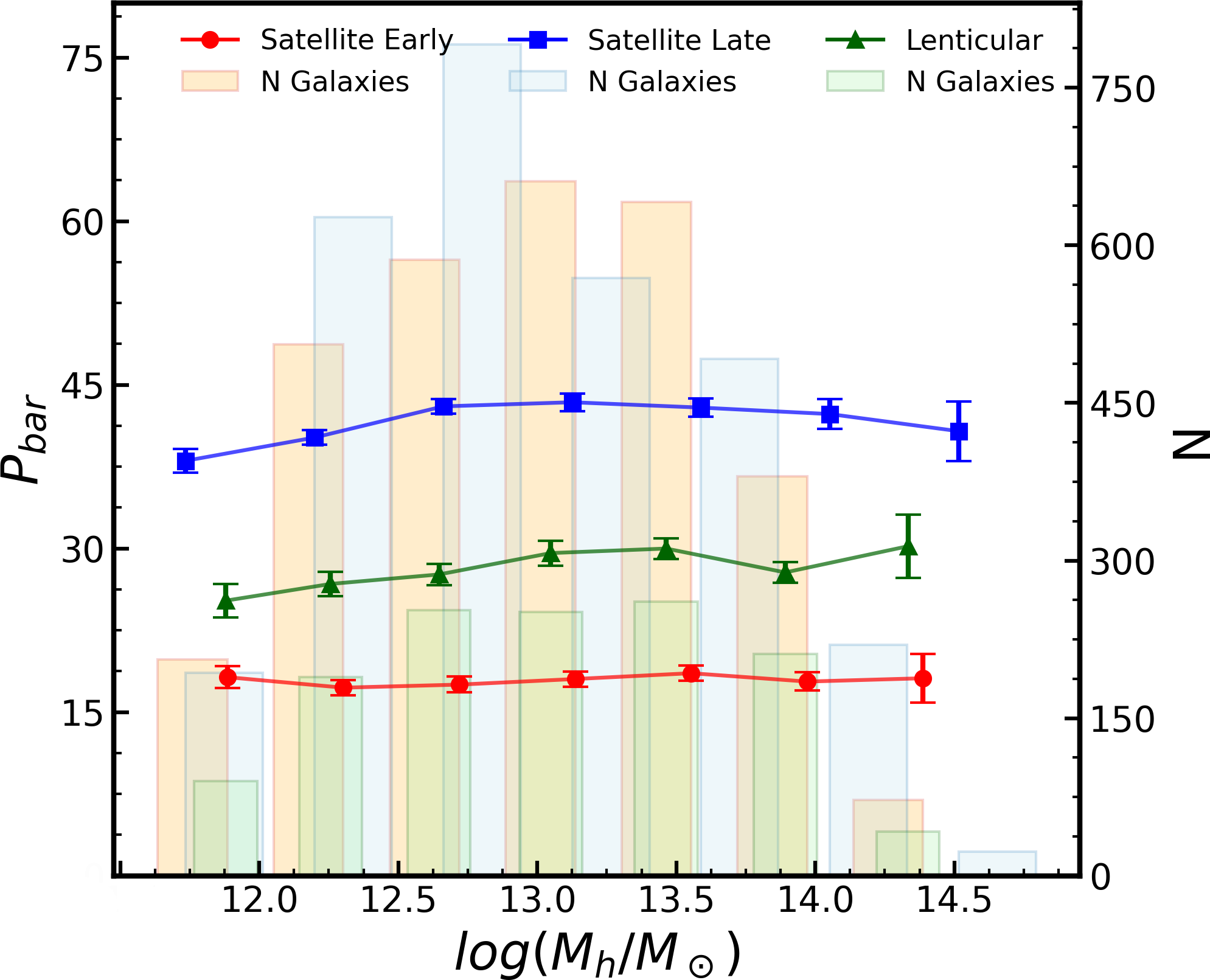}\hfill
    \includegraphics[width=0.32\textwidth]{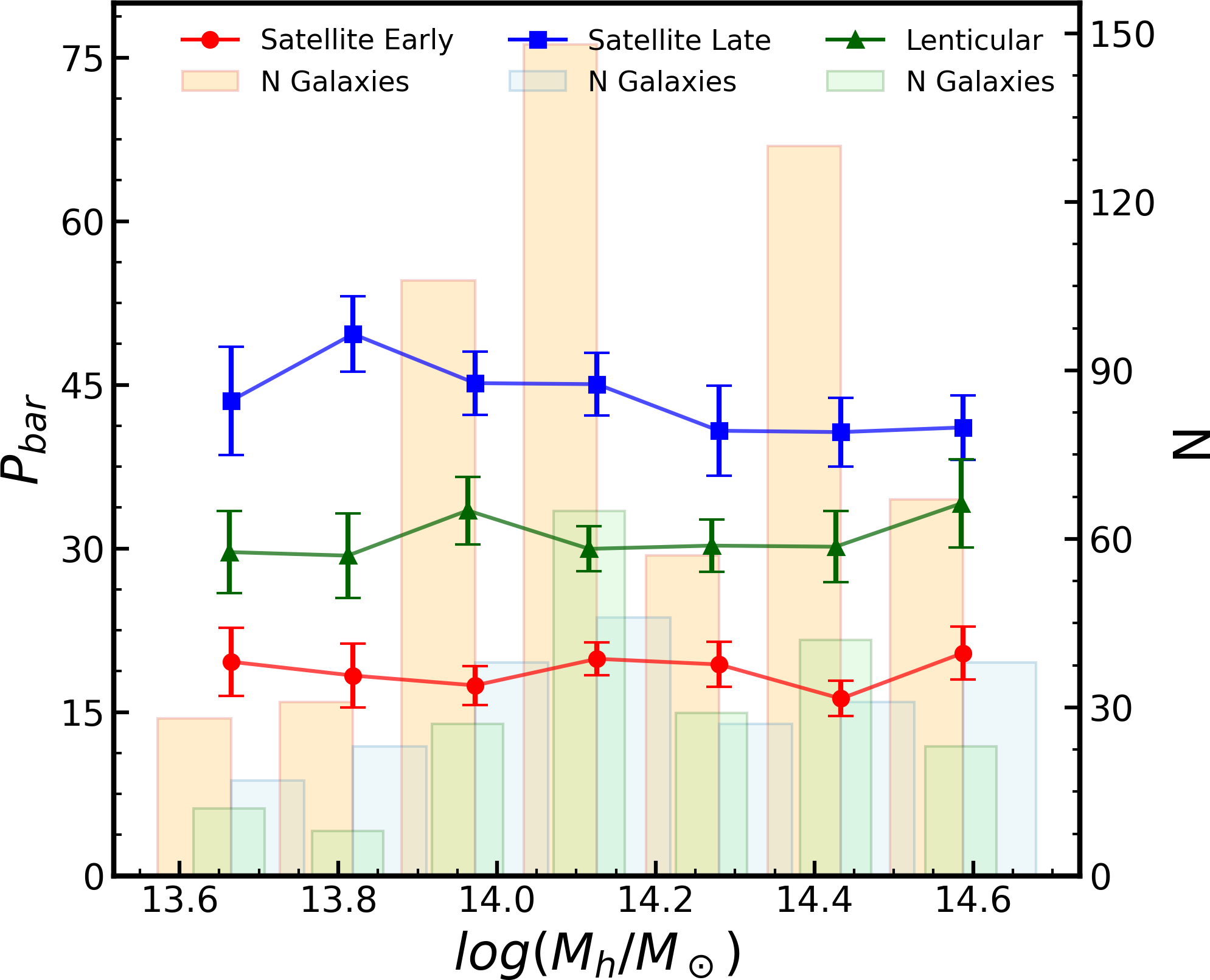}
    
    \includegraphics[width=0.32\textwidth]{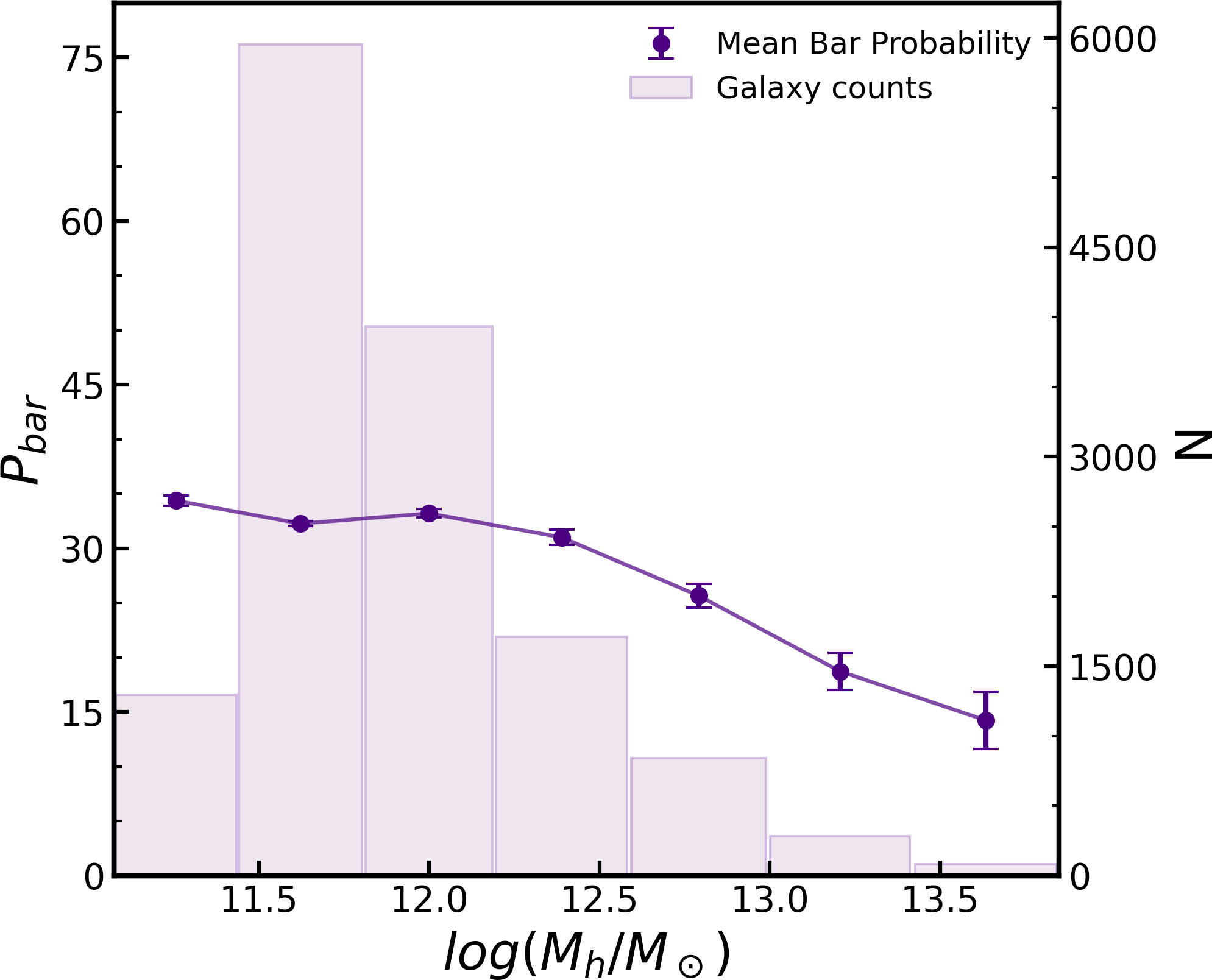}\hfill
    \includegraphics[width=0.32\textwidth]{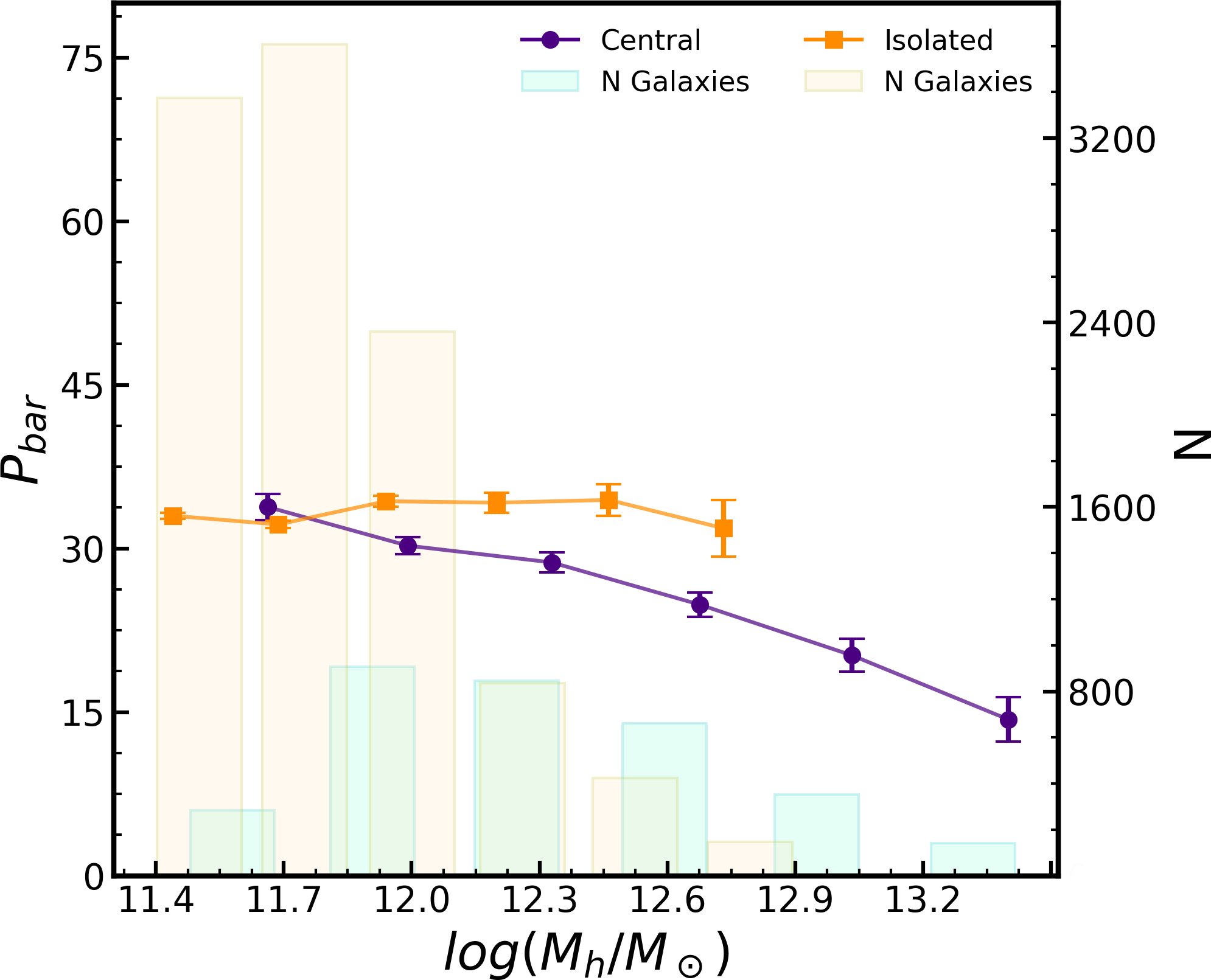}\hfill
    \includegraphics[width=0.32\textwidth]{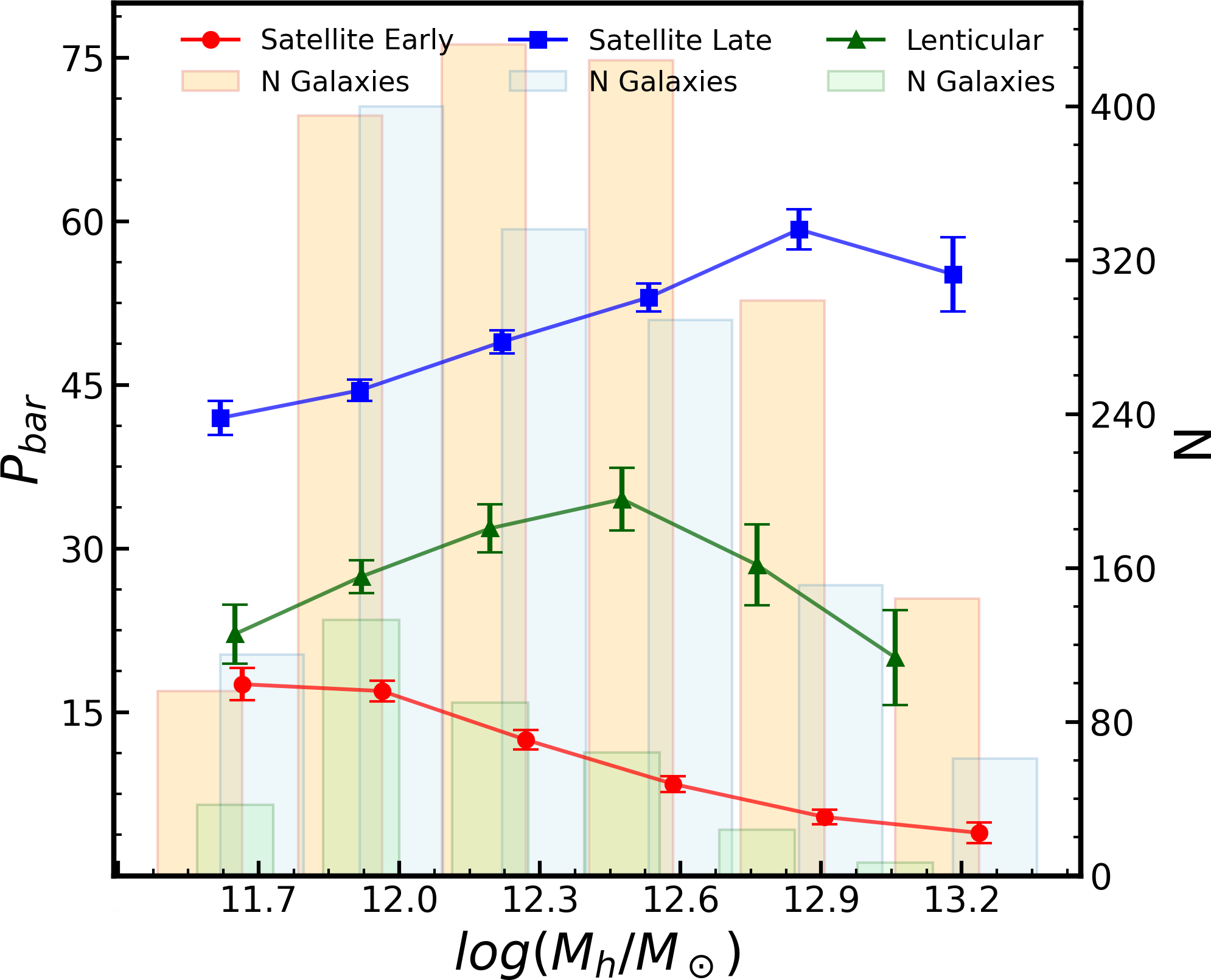}
    
    \caption{Probability of barred galaxies ($P_{\mathrm{bar}}$) as a function of halo mass $\log(M_{\mathrm{h}}/M_{\odot})$.
    Top row: full satellite sample (left panel), subsample with $N < 50$ (middle panel), and subsample with $N \geq 50$ (right panel).
    Bottom row: full central galaxy sample (left panel), centrals divided into group and isolated systems (middle panel), and subsample with $N < 50$ (right panel).
    The middle and right panels in the top row, as well as the right panel in the bottom row, display early-type, late-type, and lenticular (S0; $P(\mathrm{S0}) > 0.80$) galaxies separately.}
    
    \label{fig:fig7}
\end{figure*}

Figure~\ref{fig:fig7} presents $P_{\mathrm{bar}}$ as a function of halo mass for satellite galaxies
(top row) and central galaxies (bottom row).
For satellites, the left panel shows the full sample, while the middle and right panels separate
systems hosted by haloes with fewer than 50 members ($N < 50$) and richer haloes
($N \geq 50$), respectively.
In the latter two cases, early--type, late--type, and lenticular (S0;
$P(\mathrm{S0}) > 0.80$) galaxies are shown independently, allowing us to isolate the role of
morphology in shaping the bar--halo mass relation.

For satellite galaxies, $P_{\mathrm{bar}}$ exhibits a weak dependence on halo mass, 
remaining approximately constant across most of the explored halo mass range. 
While a mild enhancement may be present at intermediate halo masses, 
no strong monotonic trend is observed toward the most massive systems. 

When morphology is taken into account, late-type and lenticular satellites 
consistently display higher bar probabilities than early-type satellites 
at fixed halo mass, reinforcing the dominant role of internal structure. 
The relatively flat behaviour of $P_{\mathrm{bar}}$ with halo mass suggests 
that, although environmental processes such as tidal interactions, harassment, 
and ram-pressure stripping are expected to operate more efficiently in 
massive haloes \citep{Boselli+2006}, their impact on the presence of stellar 
bars appears to be secondary compared to morphology.

The bottom row of Fig.~\ref{fig:fig7} shows the corresponding trends for central galaxies. 
The left panel presents the full central population, while the middle panel distinguishes 
between group centrals and isolated centrals, and the right panel focuses on centrals in 
systems with $N < 50$ members, further split by morphological type.

In contrast to satellites, central galaxies exhibit a clearer correlation between 
$P_{\mathrm{bar}}$ and halo mass. Bar probabilities tend to increase toward more massive 
haloes, reflecting the well-known stellar mass--halo mass relation for central galaxies, 
where more massive centrals reside in more massive haloes. Since the bar likelihood 
itself depends on stellar mass, this coupling naturally induces an apparent dependence 
on halo mass.

This behaviour becomes particularly evident in the lower right panel, where morphology 
is explicitly separated. In the most massive haloes, central galaxies are predominantly 
early-type systems, which exhibit intrinsically low bar probabilities. Consequently, 
the overall decline of $P_{\mathrm{bar}}$ at the highest halo masses is largely driven 
by the morphological composition of the central population. 

\begin{figure*}
    \centering
    \includegraphics[width=0.32\textwidth]{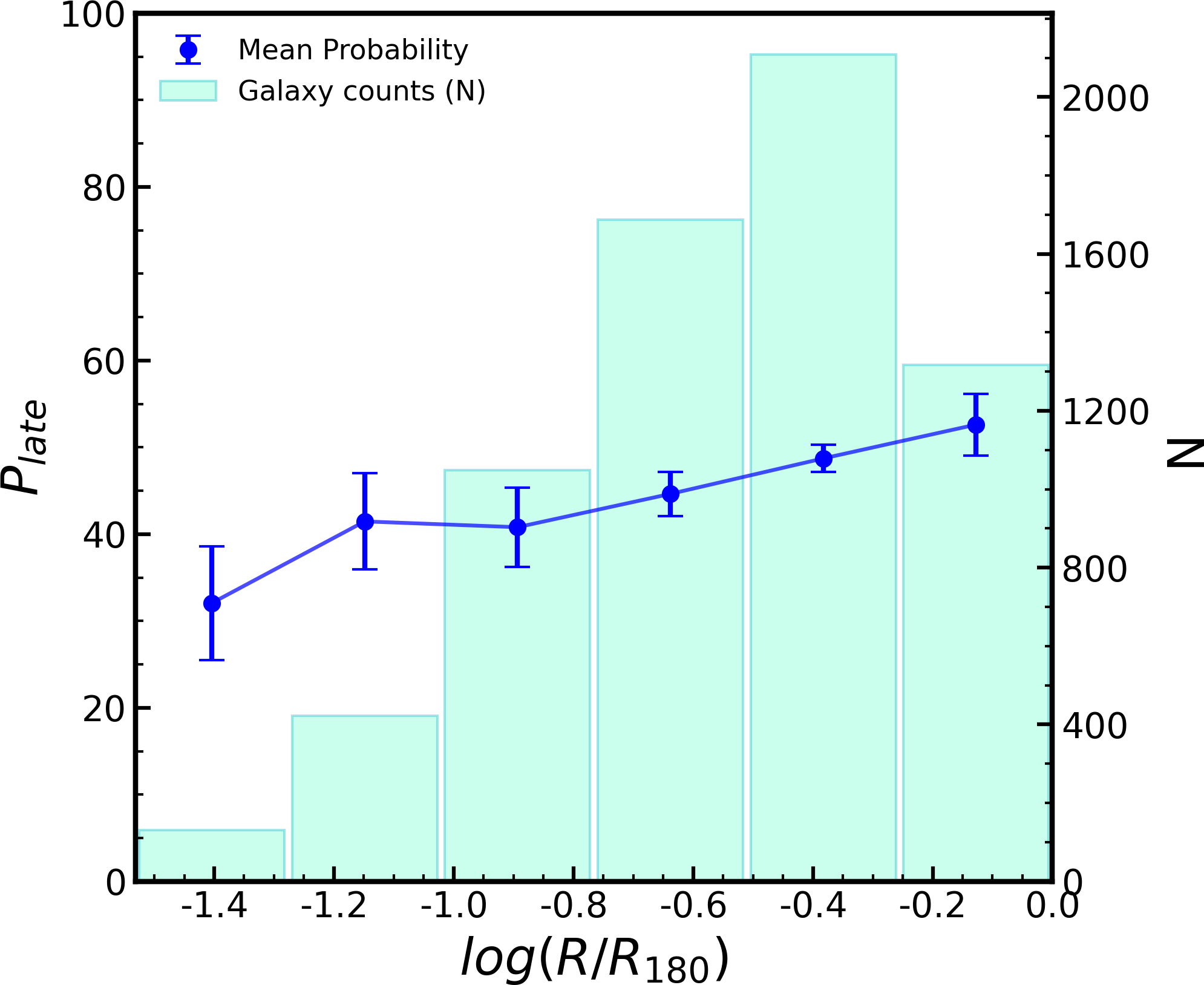}\hfill
    \includegraphics[width=0.32\textwidth]{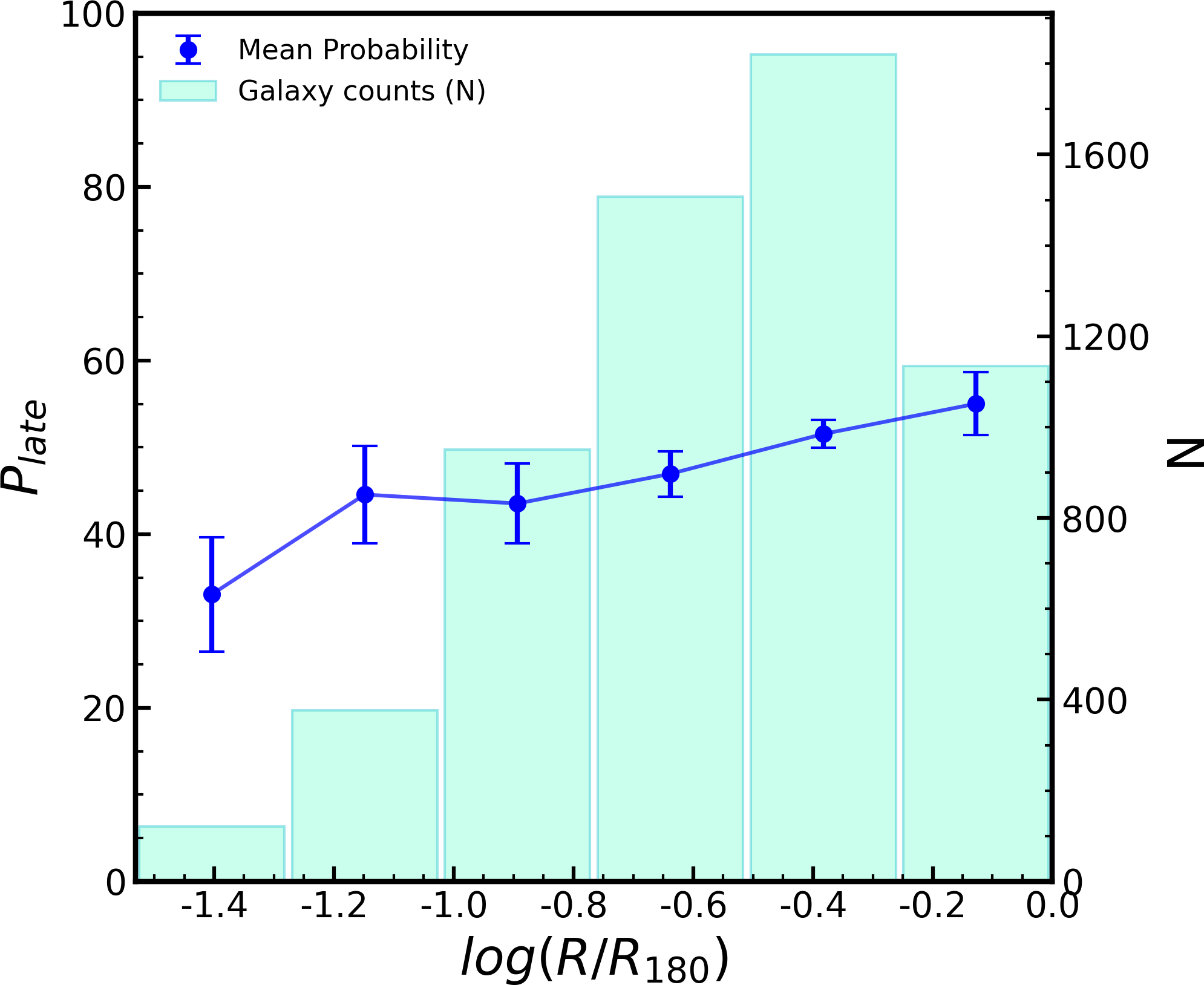}\hfill
    \includegraphics[width=0.32\textwidth]{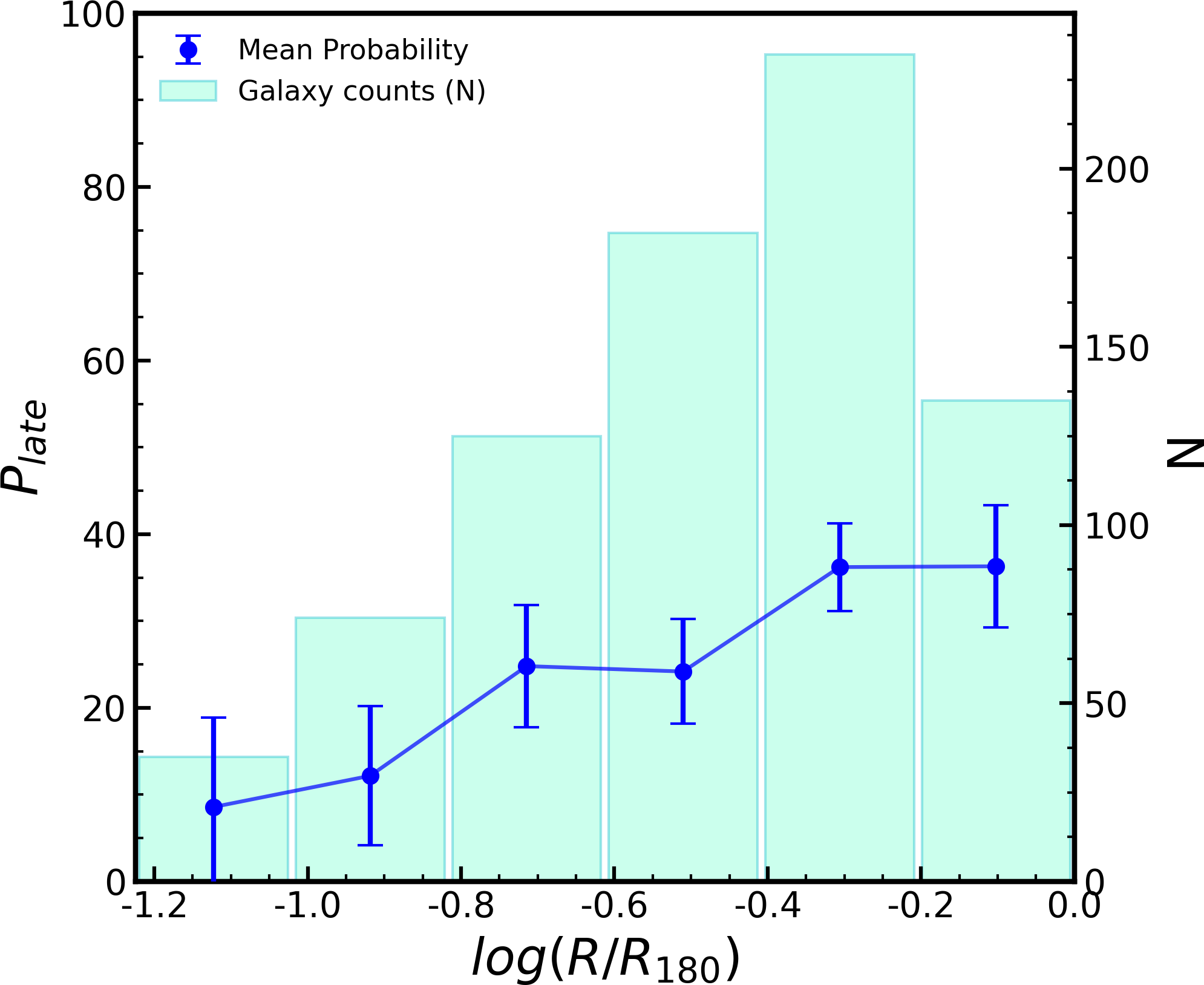}
    \caption{Probability of late-type galaxies ($P_{\mathrm{late}}$) as a function of the normalized cluster-centric distance $\log(R/R_{180})$.
    Left panel: all late-type galaxies.
    Middle panel: subsample with $N < 50$.
    Right panel: subsample with $N \geq 50$.}
    \label{fig:fig8}
\end{figure*}

\begin{figure*}
    \centering
    \includegraphics[width=0.32\textwidth]{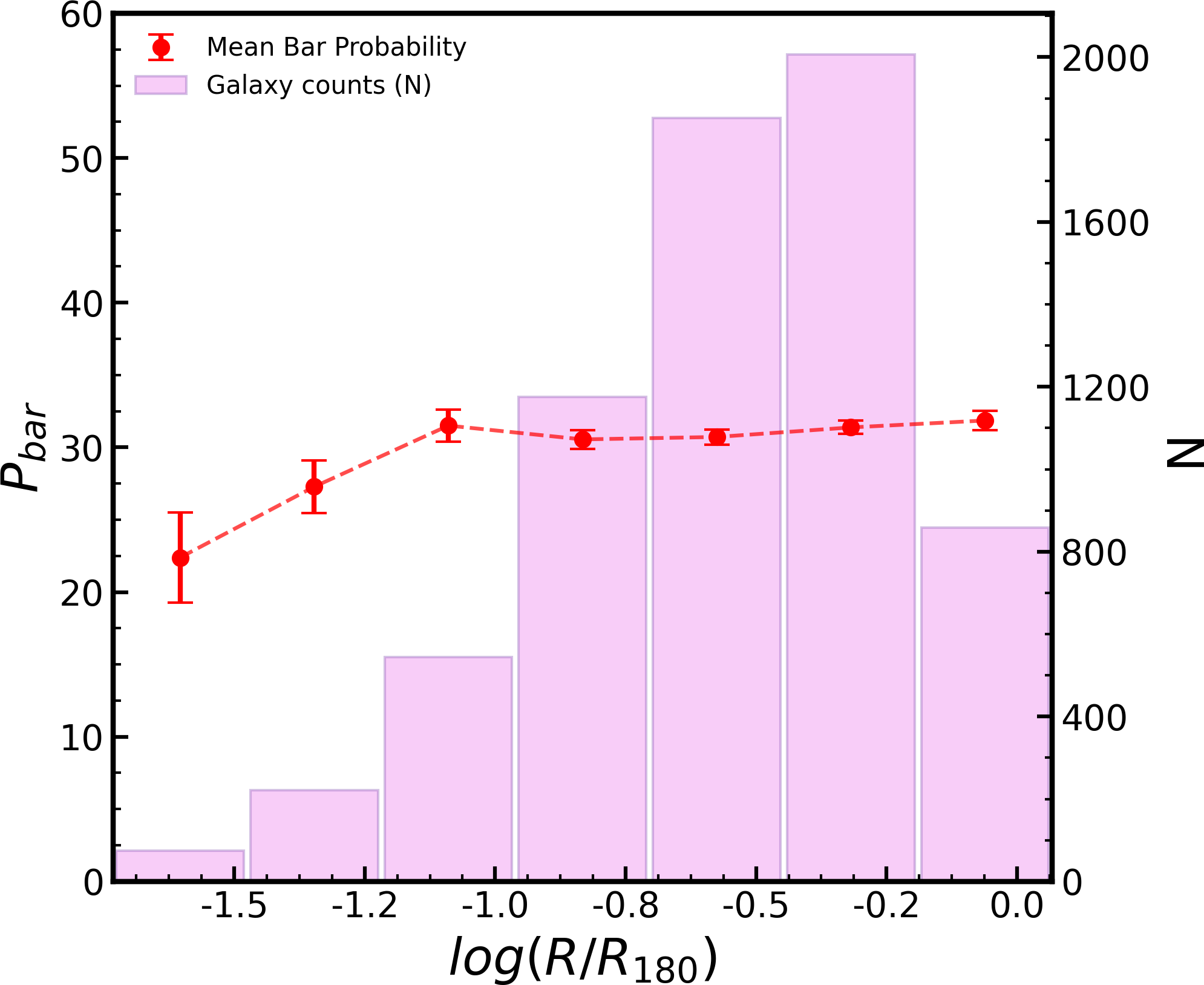}\hfill
    \includegraphics[width=0.32\textwidth]{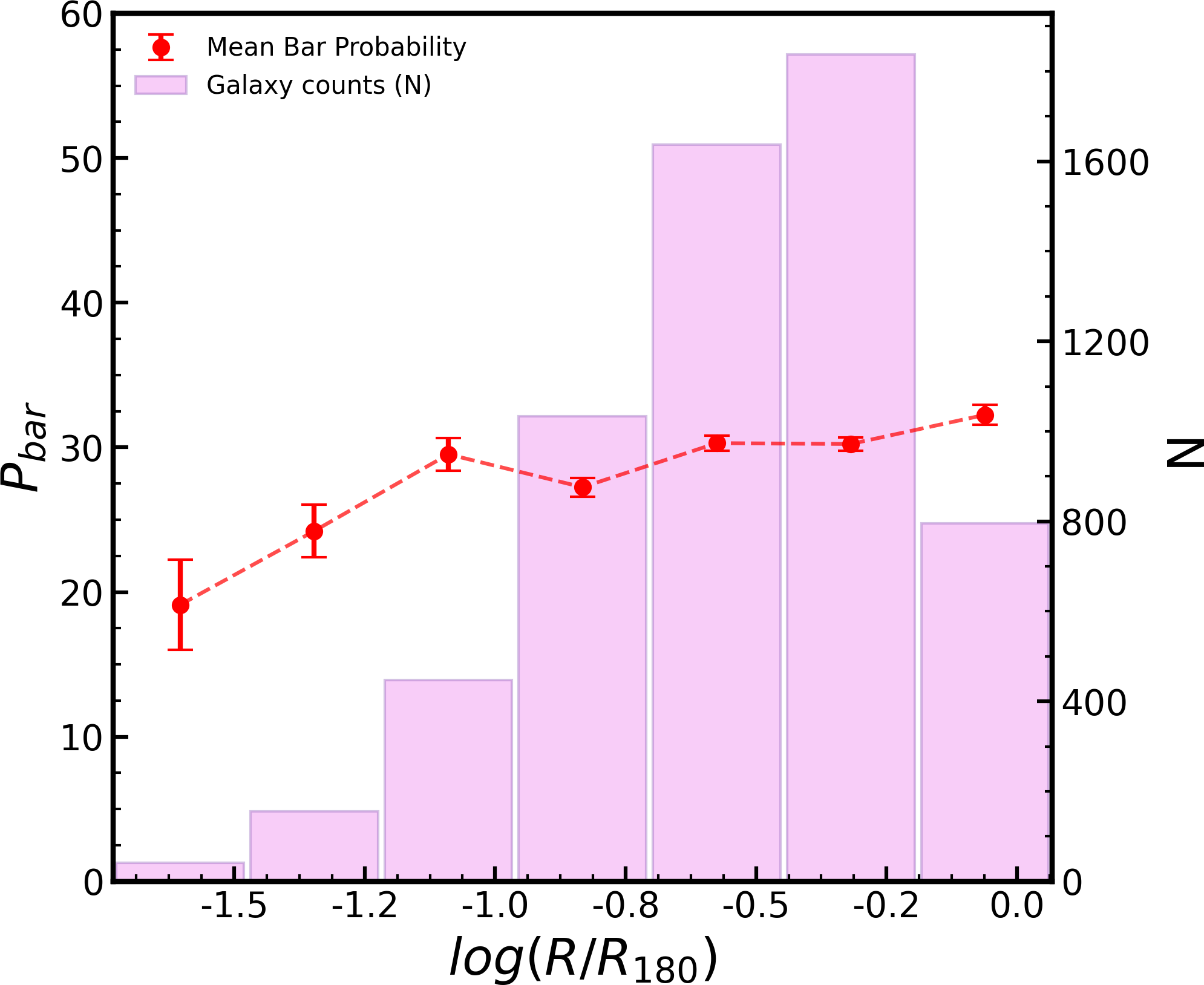}\hfill
    \includegraphics[width=0.32\textwidth]{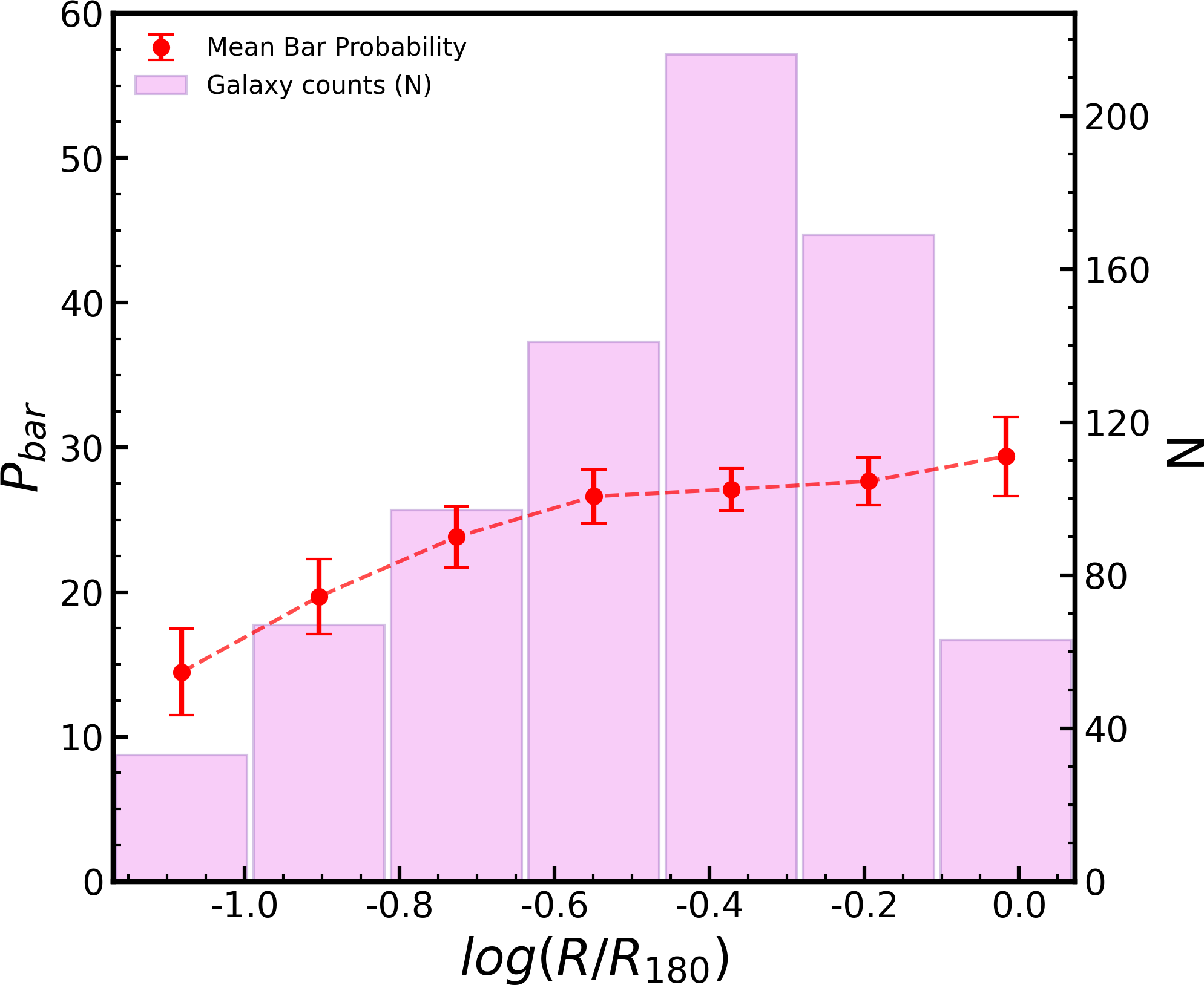}
    \caption{Probability of barred galaxies ($P_{\mathrm{bar}}$) as a function of the normalized cluster-centric distance $\log(R/R_{180})$.
    Left panel: all satellite galaxies.
    Middle panel: subsample with $N < 50$.
    Right panel: subsample with $N \geq 50$.}
    \label{fig:fig9}
\end{figure*}

At fixed morphology, however, a different trend emerges: late-type central galaxies 
show increasing bar probabilities toward higher halo masses, consistent with the 
correlation between stellar mass and halo mass for centrals. In contrast, early-type 
centrals maintain systematically low $P_{\mathrm{bar}}$ values across the halo mass range. 
These trends reinforce that morphology — and its connection to stellar mass —
is the strongest observed correlate of the halo mass dependence of
$P_{\mathrm{bar}}$ in our sample.

The interpretation of these trends is consistent with the broader theoretical
picture of bar formation and evolution established by numerical simulations.
Classical $N$-body experiments demonstrated that self-gravitating stellar
discs can develop large-scale bar instabilities \citep{Hohl+1971}, while
subsequent studies showed that the evolution of bars is strongly coupled to
the redistribution of angular momentum between the stellar disc, outer
galaxy, and dark matter halo \citep{Combes+1981,
Athanassoula+2003}. In this framework, the halo does not necessarily act as
a simple external parameter controlling whether a bar forms. Rather, the
dynamical structure and response of the halo can influence the growth,
strength, and subsequent evolution of an already unstable stellar disc.

Recent idealized and cosmological simulations have extended this picture
by demonstrating that the interaction between the stellar disc and its
surrounding halo can affect both the formation timescale and the subsequent
evolution of bars \citep{Frosst+2024,Fragkoudi+2025}. In
particular, simulations indicate that bar formation is closely related to
the internal dynamical state of the disc, including its stellar mass
fraction, stability, and structural properties, while the response of a
live dark matter halo can modify the growth and evolution of the bar.
Recent cosmological simulations further suggest that bars may arise through
different evolutionary channels, including secular disc instabilities and
tidally induced processes \citep{Frosst+2026}. These results provide a
physical context for our finding that the observed halo-mass dependence of
$P_{\mathrm{bar}}$ is closely coupled to stellar mass and morphology, and
support a distinction between the conditions associated with bar formation
and the subsequent environmental evolution of bars.

Isolated centrals display slightly higher bar probabilities than group centrals at fixed 
halo mass, in agreement with previous studies reporting a reduced environmental impact 
on bar formation for galaxies evolving in relative isolation 
\citep{Cervantes-Sodi+2015}.

The comparison between satellite and central populations indicates that we find no strong evidence that halo-scale environment, as traced by halo mass, directly regulates bar formation. Instead, its influence appears to be largely indirect,
acting through its role in driving morphological transformation, stellar mass assembly,
and disc stability.

For central galaxies, the apparent dependence of $P_{\mathrm{bar}}$ on halo mass is primarily a consequence of the tight stellar mass--halo mass relation, whereby more massive centrals reside in more massive haloes. In contrast, satellite galaxies show only a weak variation of $P_{\mathrm{bar}}$ with halo mass, consistent with the idea that they do not dominate the halo potential and that their bar likelihood is mainly governed by their internal structural properties.

\subsection{Radial trends in bar likelihood and the morphology–density relation}

\begin{figure*}
    \centering
    \includegraphics[width=0.32\textwidth]{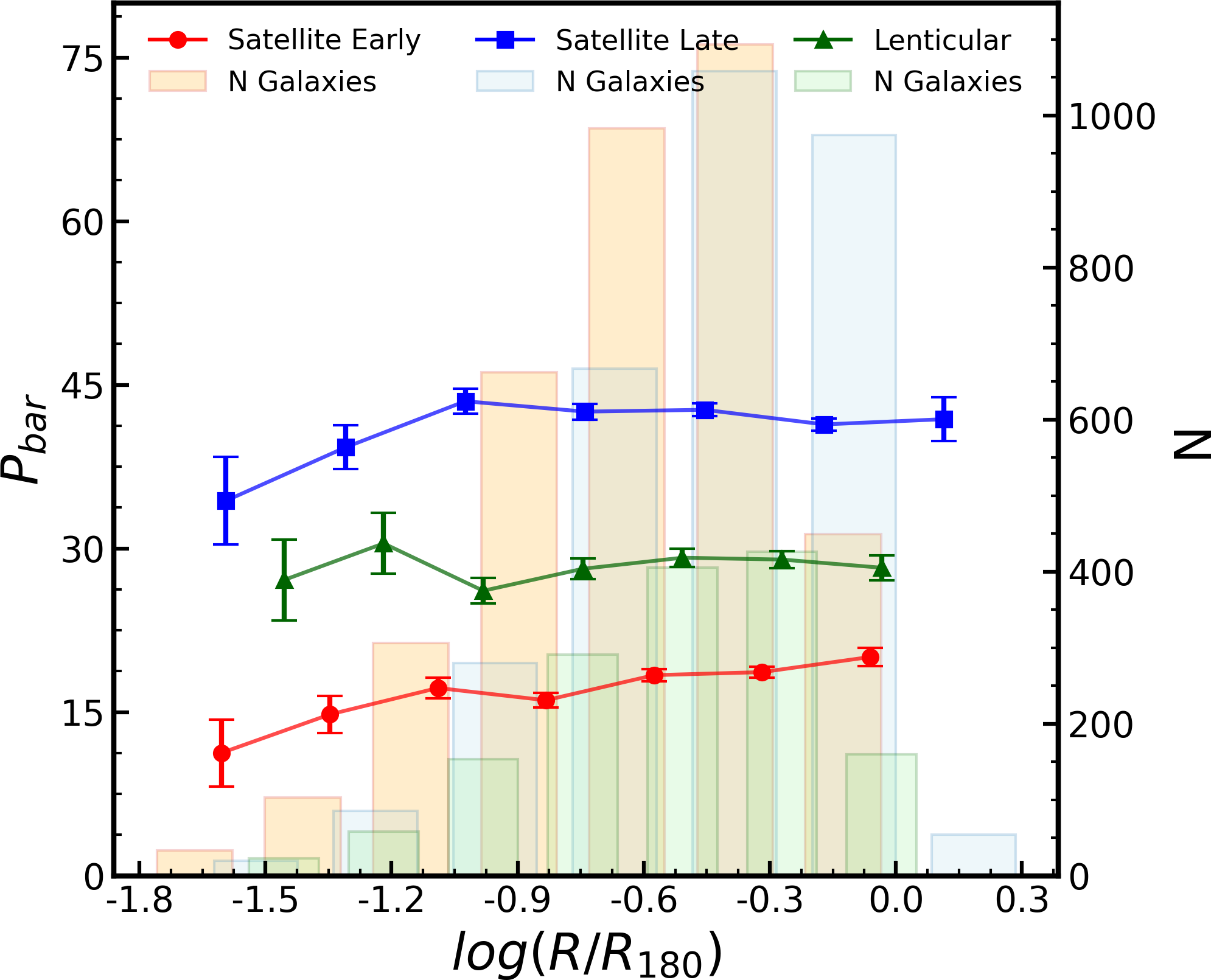}\hfill
    \includegraphics[width=0.32\textwidth]{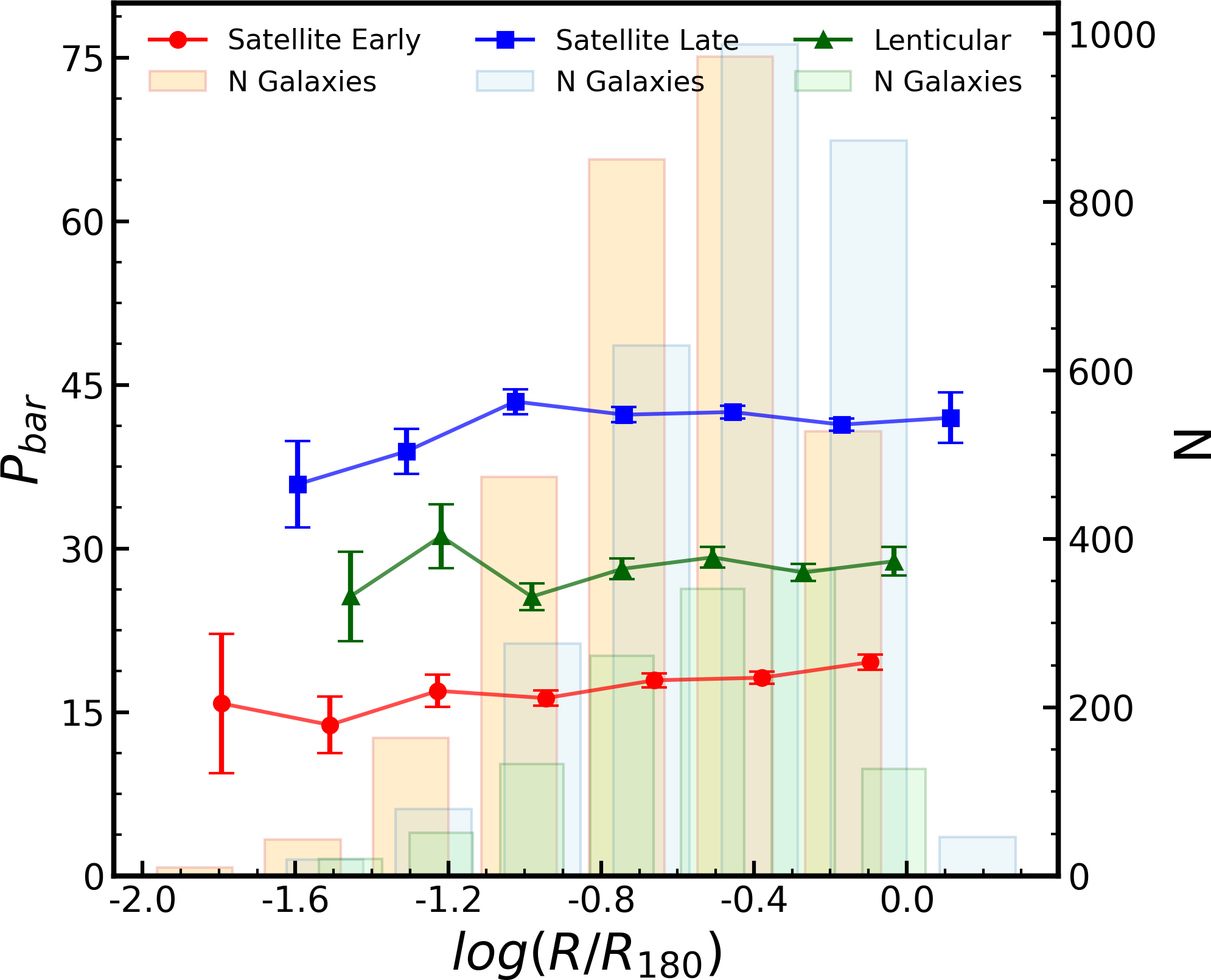}\hfill
    \includegraphics[width=0.32\textwidth]{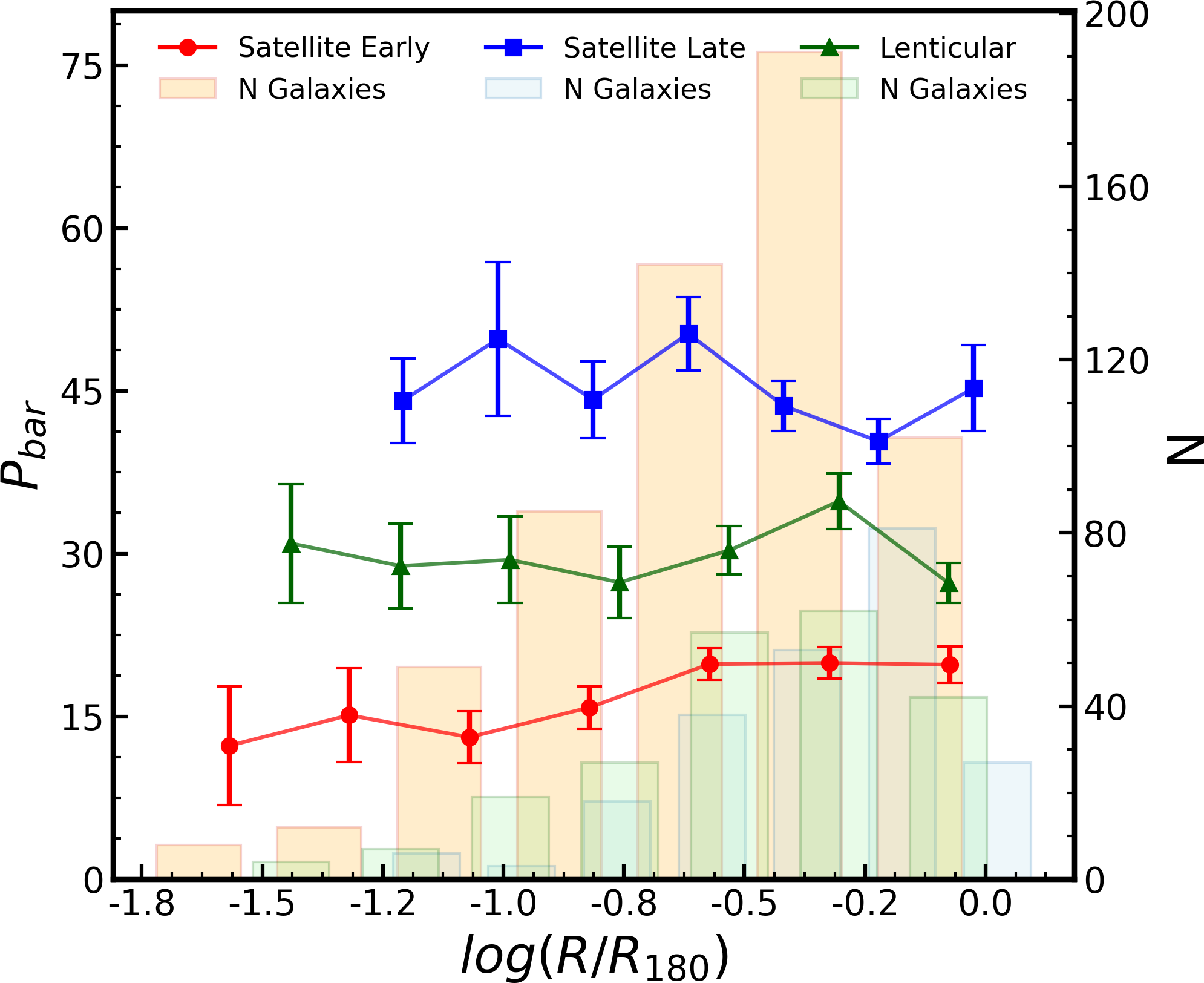}
    \caption{Probability of barred galaxies ($P_{\mathrm{bar}}$) as a function of the normalized cluster-centric distance $\log(R/R_{180})$.
    Left panel: all satellite galaxies.
    Middle panel: satellite galaxies in systems with $N < 50$.
    Right panel: satellite galaxies in systems with $N \geq 50$.}
    \label{fig:fig10}
\end{figure*}

Given that our previous results indicate that  the bar likelihood is primarily governed 
by morphology, any apparent radial dependence of bar probability is expected 
to arise indirectly through changes in the morphological mix, rather than from 
a direct environmental effect.

Figure~\ref{fig:fig8} shows the probability of a satellite galaxy being late–type 
($P_{\mathrm{late}}$) as a function of the normalized cluster–centric distance, 
expressed as $\log(R/R_{180})$. Only satellite galaxies are considered in this analysis, 
ensuring that the trends reflect environmental effects within host haloes rather than 
properties of central systems.

The left panel presents the full satellite population, while the middle and right panels 
separate systems with $N < 50$ and $N \geq 50$ members, respectively. 

A clear radial trend is observed in all cases: the probability of a galaxy being late-type 
increases with cluster-centric distance. In the full sample, $P_{\mathrm{late}}$ rises 
moderately from approximately 40\% in the central regions to about 55\% toward the outskirts. 
A similar behaviour is found in systems with $N < 50$ members.

In richer systems ($N \geq 50$), the radial dependence becomes more pronounced. 
The innermost regions are strongly dominated by early-type satellites, with a very low 
late-type probability, while $P_{\mathrm{late}}$ increases toward larger radii, 
reaching values of approximately 40\% in the outer parts. This behaviour reflects 
the well-established morphology–density relation operating within cluster environments.

Figure~\ref{fig:fig9} displays the probability of hosting a bar 
($P_{\mathrm{bar}}$) as a function of the normalized cluster–centric distance, 
expressed as $\log(R/R_{180})$, for satellite galaxies. The left panel shows 
the full satellite sample, while the middle and right panels correspond to 
systems with $N < 50$ and $N \geq 50$ members, respectively.

In the full satellite population (left panel), $P_{\mathrm{bar}}$ decreases 
toward the cluster centre, reaching values of approximately 30\%, and 
remains relatively constant at $\sim$40\% in the outer regions. 
A similar but more pronounced trend is observed in poorer systems 
($N < 50$), where the bar probability drops to $\sim$18\% in the inner 
regions and increases to $\sim$30\% at larger radii, remaining roughly 
flat beyond intermediate distances.

In richer systems ($N \geq 50$), the central suppression becomes even 
stronger, with $P_{\mathrm{bar}}$ decreasing to $\sim$14\% near the cluster 
core and rising to about 27\% toward the outskirts.

However, when compared directly with Fig.~\ref{fig:fig8}, 
the radial behaviour of $P_{\mathrm{bar}}$ closely mirrors 
the variation in morphological composition. Since the cluster cores are dominated 
by early-type satellites — which intrinsically exhibit low bar 
probabilities — the central decrease in $P_{\mathrm{bar}}$ primarily 
reflects the changing morphological mix rather than a direct 
environmental suppression of bars. 

Therefore, the observed radial trend appears to be largely explained by the morphology--density relation operating within clusters. The results do not provide strong evidence that environment directly regulates the probability of hosting a bar. Nevertheless, weaker residual environmental effects cannot be excluded from the trends shown here.

Figure~\ref{fig:fig10} shows the probability of hosting a bar 
($P_{\mathrm{bar}}$) as a function of the normalized cluster–centric 
distance ($R/R_{180}$) for different morphological subsamples of 
satellite galaxies. The left panel presents the full satellite population, 
while the middle and right panels correspond to systems with $N < 50$ 
and $N \geq 50$ members, respectively.

In the full satellite sample (left panel), late-type galaxies exhibit 
the highest bar probabilities, reaching values of approximately 40\% 
in the outer regions and decreasing toward the cluster centre. 
Lenticular (S0) galaxies show intermediate bar probabilities, around 
$\sim$28\%, with a relatively weak dependence on cluster-centric 
distance. Early-type satellites display the lowest bar likelihood, 
remaining below $\sim$12\%, and also show a decline toward the inner 
cluster regions.

A similar behaviour is observed in poorer systems ($N < 50$; middle panel), 
where late-type galaxies again present the highest bar probabilities, 
decreasing toward the cluster core, while lenticular systems remain 
approximately constant with radius and early-type galaxies maintain 
systematically low bar fractions.

In richer systems ($N \geq 50$; right panel), late-type satellites show 
an approximately constant bar probability with radius, although their 
number becomes progressively smaller toward the cluster centre. 
Lenticular galaxies exhibit a similarly flat behaviour. In contrast, 
early-type satellites show a clear decrease in $P_{\mathrm{bar}}$ toward 
the innermost regions.

Overall, the dominant segregation in bar probability is morphological: 
late-type satellites are consistently more likely to host bars than 
lenticular systems, which in turn exhibit higher probabilities than 
early-type galaxies. The radial trends observed within each richness 
regime are secondary and generally modest compared to the strong 
dependence on morphology.

\begin{figure}
    \centering
    \includegraphics[width=\columnwidth]{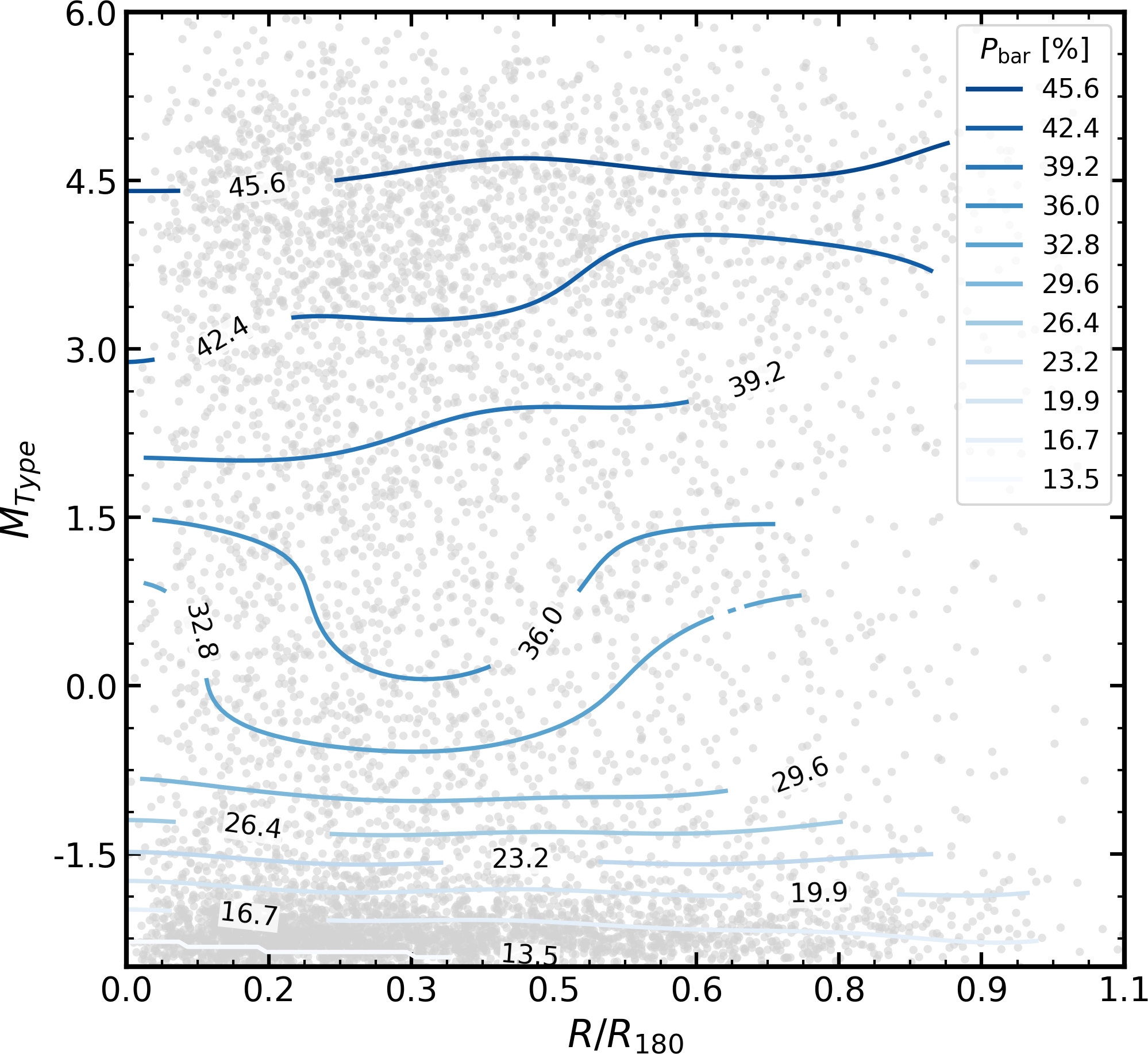}
    \caption{Probability of barred galaxies ($P_{\mathrm{bar}}$) as a function of the normalized cluster-centric distance $\log(R/R_{180})$.
    The blue isocontours indicate the regions of highest probability of hosting a bar.}
    \label{fig:fig11}
\end{figure}

Figure~\ref{fig:fig11} presents the probability of hosting a bar 
($P_{\mathrm{bar}}$) as a function of morphological type and normalized 
cluster–centric distance ($R/R_{180}$) for satellite galaxies. 
The blue isocontours trace regions of equal bar probability, providing 
a two-dimensional view of how $P_{\mathrm{bar}}$ varies across both 
morphology and cluster environment.

A strong morphological dependence is immediately apparent. 
Late-type galaxies (with $M_{\rm type} \gtrsim 4$) exhibit the highest 
bar probabilities, reaching values of $\sim$45\%, which remain nearly 
constant across the full radial range. At intermediate morphological 
types ($M_{\rm type} \sim 3$), the bar probability is slightly lower 
($\sim$42\%) and shows a mild increase toward larger cluster-centric 
distances. A similar, though weaker, radial enhancement is observed 
around $M_{\rm type} \sim 1.5$, where $P_{\mathrm{bar}}$ reaches 
$\sim$40\% in the outer regions.
For earlier morphological types ($M_{\rm type} \lesssim 1.5$), 
the behaviour changes. The isocontours show a shallow U-shaped pattern, 
with $P_{\mathrm{bar}}$ reaching approximately 36\% at intermediate 
radii ($R/R_{180} \sim 0.3$), while decreasing both toward the cluster centre and toward earlier types. In the earliest systems 
($M_{\rm type} \lesssim -1$), the bar probability drops to 
$\sim$26\% and continues to decline steadily, largely independent 
of cluster-centric distance.

Overall, the isocontour structure confirms that the dominant driver 
of bar probability is morphology. The contours are predominantly 
horizontal, indicating that variations with cluster-centric distance 
are modest compared to the strong systematic change across 
morphological type. Any radial modulation is therefore secondary and 
appears to reflect the changing morphological mix within clusters 
rather than providing evidence for a direct environmental regulation 
of bar formation.

A quantitative residual analysis controlling simultaneously for stellar mass 
and morphology is presented in next Section~\ref{subsec:residual_environment}. 
This analysis provides an additional test of whether the environmental trends 
described above persist after accounting for the changing stellar-mass and 
morphological composition of the satellite population.

\subsection{Residual environmental dependence of the bar likelihood}
\label{subsec:residual_environment}

The trends discussed above show that bar likelihood varies strongly with
galaxy morphology and stellar mass, while the apparent dependence on halo
mass and cluster-centric distance is closely connected to the changing
composition of the galaxy population. To quantify whether an additional
environmental dependence remains after accounting for these properties, we
perform a residual analysis using stellar mass and morphology as control
variables.

For this analysis, the satellite population is divided into four mutually
exclusive morphological classes: S0, early-type non-S0, late-I
($0<T\leq3$), and late-II ($T>3$). Each morphological class is further
divided into four stellar-mass bins, resulting in 16 morphology--stellar-mass
reference cells (see Table~\ref{tab:expected_pbar}). The expected bar
likelihood, $P_{\mathrm{bar,exp}}$, is calculated as the mean bar likelihood
of the satellite galaxies within each reference cell. The residual bar
likelihood is defined as
\begin{equation}
\Delta P_{\mathrm{bar}} =
P_{\mathrm{bar,obs}} - P_{\mathrm{bar,exp}},
\label{eq:delta_pbar}
\end{equation}
where $P_{\mathrm{bar,obs}}$ is the observed mean bar likelihood in a given
environmental bin. For each environmental bin, $P_{\mathrm{bar,exp}}$ is
obtained by averaging the reference values corresponding to the
morphology--stellar-mass class of the galaxies in that bin. A residual of
zero therefore indicates consistency with the expectation based on the
stellar-mass and morphological composition of the reference sample. The
reference values are summarized in Appendix~\ref{app:residual_environment}.

\begin{figure}
    \centering
    \includegraphics[width=\columnwidth]{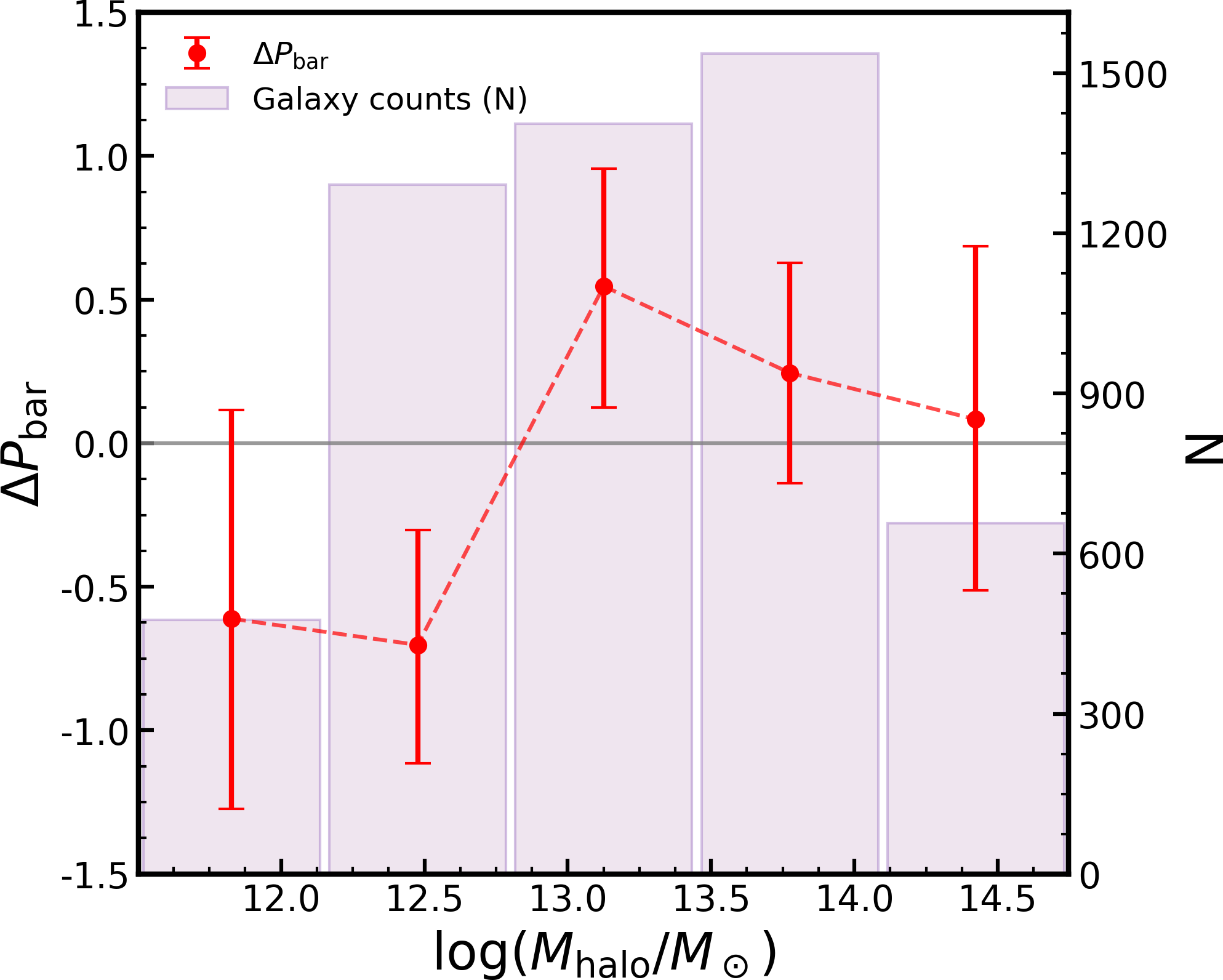}
    \includegraphics[width=\columnwidth]{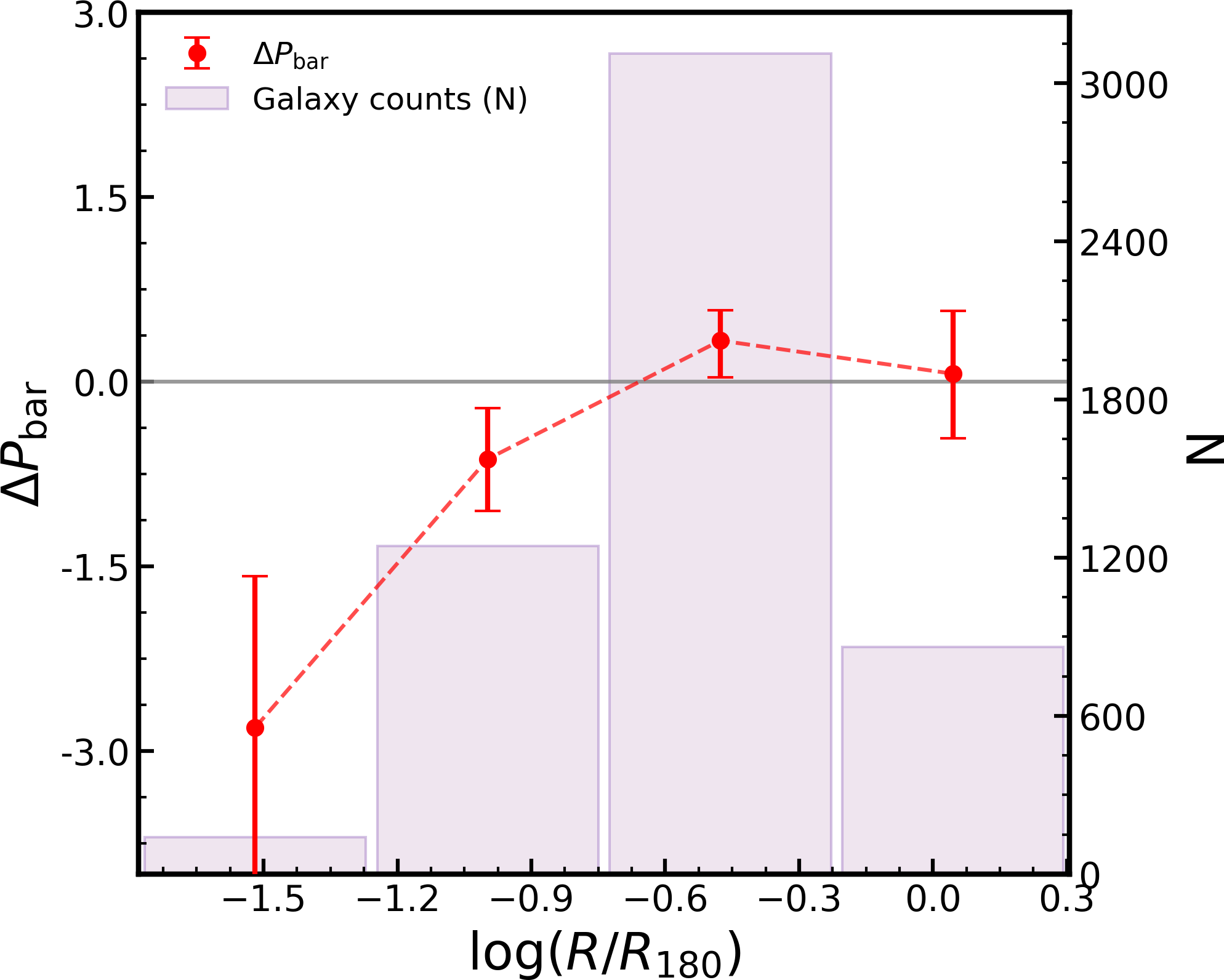}
    \caption{Residual bar likelihood after controlling for stellar mass and
    morphology for the satellite population. The upper panel shows
    $\Delta P_{\rm bar}$ as a function of halo mass, while the lower panel
    shows the corresponding residual as a function of normalized
    cluster-centric radius, $R/R_{180}$. The points show the mean residual
    in each environmental bin, with bootstrap uncertainties. The histograms
    indicate the number of galaxies contributing to each bin. The horizontal
    line at $\Delta P_{\rm bar}=0$ represents the expectation obtained from
    the stellar-mass and morphological control. The residuals remain broadly
    consistent with zero, indicating no statistically significant systematic
    environmental dependence within the sensitivity of the present
    analysis.}
    \label{fig:residual_environment}
\end{figure}

Figure~\ref{fig:residual_environment} shows the resulting residuals as a
function of halo mass and normalized cluster-centric distance. The residuals
remain broadly consistent with zero across the explored halo-mass range,
with no systematic monotonic trend. A similarly weak pattern is found as a
function of normalized cluster-centric distance for satellite galaxies. The
smallest radial bin is excluded because it contains only nine galaxies and
therefore does not provide sufficient statistical sampling for a robust
estimate.

These results provide a quantitative test of the interpretation developed
in the previous sections. After accounting simultaneously for stellar mass
and morphology, we find no statistically significant residual dependence of
$P_{\mathrm{bar}}$ on halo mass or normalized cluster-centric distance within
the sensitivity of the present sample. This result does not demonstrate that
environment has no physical influence on bar formation; rather, it indicates
that any additional halo-scale environmental contribution remaining after
controlling for stellar mass and morphology is not statistically resolved in
the present analysis. Weaker environmental effects below this statistical
sensitivity cannot be excluded.

\section{Summary and Conclusions}\label{sec:conclusion}

In this work, we have investigated how the probability of hosting a stellar bar 
depends on both intrinsic galaxy properties and environment. Using a 
volume-limited sample of central and satellite galaxies 
spanning a broad range of environments, from isolated systems through
low- and high-richness groups to massive haloes, drawn from the 
NASA--Sloan Atlas, complemented with halo-based environmental information and 
probabilistic morphological classifications, we analysed the behaviour of the 
bar likelihood as a function of stellar mass, colour, morphology, halo mass, 
and cluster-centric distance. By adopting a fully probabilistic approach to bar identification, we move beyond traditional binary classifications and quantify $P_{\mathrm{bar}}$ as a continuous variable. This allows us to disentangle the relative roles of stellar mass, morphology, and environment in a self-consistent framework, extending previous studies to a halo-based description of environment.

Our results show that stellar mass and galaxy morphology are jointly
associated with the probability of hosting a stellar bar. At fixed morphological type, stellar mass 
plays a secondary role, with late-type galaxies showing an increase of bar 
likelihood toward higher masses, while early-type systems maintain low bar 
probabilities largely independent of stellar mass.

Clear trends are observed between bar likelihood and environment, including 
halo mass and cluster-centric distance. However, these dependencies are largely 
driven by variations in the underlying morphological composition of the galaxy 
population. This reflects the morphology--density relation, whereby late-type 
galaxies are more frequent in low-density regions and at large cluster-centric 
distances, while early-type systems dominate dense environments and cluster 
cores. As a result, the observed decrease in $P_{\mathrm{bar}}$ toward the 
centres of massive haloes primarily reflects the increasing fraction of 
early-type galaxies, rather than a direct environmental suppression of bars.

The role of halo mass is consistent with this interpretation. Satellite galaxies 
show only a weak dependence of $P_{\mathrm{bar}}$ on halo mass once morphology 
is taken into account. For central galaxies, an apparent increase of bar 
probability with halo mass is observed, but this trend is largely driven by the 
underlying stellar mass--halo mass relation and its coupling with morphology.
The residual analysis described in Section~\ref{subsec:residual_environment}
provides a quantitative test of this interpretation.

These results indicate that we find no strong evidence that halo-scale environment,
as traced by halo mass, directly regulates bar formation once the changing
morphological and stellar-mass composition of the galaxy population is taken into
account. The quantitative residual analysis presented in
Section~\ref{subsec:residual_environment} shows no statistically significant residual
dependence of $P_{\mathrm{bar}}$ on either halo mass or normalized cluster-centric
radius within the sensitivity of the present sample.
This result does not imply that environment has no physical influence on bar formation,
but rather that any environmental contribution remaining after accounting for stellar
mass and morphology is not statistically significant within the sensitivity of the
present sample. Weaker environmental effects below this statistical sensitivity
therefore cannot be excluded.

We note, however, that our analysis focuses on halo-scale environmental indicators.
Therefore, we cannot exclude the possibility that small-scale processes, such as
close galaxy interactions or minor mergers, may still play a role in triggering or
suppressing bar formation, as suggested by previous studies
\citep{Lee+2012, Tawfeek+2022, Li2026}.

Our observational results are also broadly consistent with recent numerical simulations. 
Cosmological simulations indicate that stellar bars form predominantly through secular 
instabilities in dynamically cold stellar discs, while environmental processes such as 
tidal encounters mainly influence the evolution, strength, and lifetime of pre-existing 
bars, and only in a minority of cases act as the primary trigger of bar formation 
(e.g. \citealt{Zana18}; \citealt{Frosst26}). 

Within this framework, the absence of a statistically significant residual
environmental dependence in our sample can be understood if galaxy morphology
is regarded as an observational proxy for the underlying dynamical state of
galaxies, including properties such as the bulge-to-disc ratio, disc stability,
gas content, and assembly history, which ultimately regulate the formation and
long-term evolution of stellar bars
(\citealt{Rosas-Guevara22}; \citealt{Lopez24}; \citealt{Lu25};
\citealt{Chim25}; \citealt{Chim26}).

Taken together, our results suggest that the commonly reported dependence
of bar fraction on environment does not necessarily reflect a direct environmental
effect on bar formation. The absence of a statistically significant residual
dependence after accounting for stellar mass and morphology is consistent with a
scenario in which environmental trends in the observed bar likelihood arise
primarily through changes in the intrinsic properties and morphological composition
of galaxies. An important aspect of this result is that it is obtained across a broad
range of halo masses and environmental regimes, with central and satellite
galaxies analysed separately, rather than from a simple cluster--field
comparison.

In this picture, galaxy morphology is not the fundamental physical
driver of bar formation, but rather an observable manifestation of the underlying
dynamical state of galaxies. We emphasize, however, that weaker environmental
effects below the statistical sensitivity of the present analysis cannot be excluded.

\section*{Acknowledgements}

The authors thank the anonymous referees for useful comments that
helped to improve the quality of the paper and clarify our results. We gratefully acknowledge financial support from the Universidad Tecnol\'ogica de Panam\'a through the Bar Detection in Galaxies (BADEG) Project. MAC and BCS acknowledge the financial support provided by PAPIIT projects IN108323 and IN111825 from DGAPA-UNAM. YLJ acknowledges support from the Agencia Nacional de Investigaci\'on y Desarrollo (ANID) through Basal project FB210003, FONDECYT Regular projects 1241426 and 123044, and  Millennium  Science Initiative Program NCN2024\_112. This study is built upon data obtained from the Sloan Digital Sky Survey (SDSS). Funding for the
SDSS and SDSS-II has been provided by the Alfred P. Sloan Foundation and Participating Institutions, the National Science Foundation, the U.S. Department of Energy, the National
Aeronautics and Space Administration, the Japanese Monbukagakusho, the Max Planck Society, and the Higher Education Funding Council for England. The SDSS web site is \url{www.sdss.org}. The SDSS is managed by the Astrophysical Research Consortium for the Participating Institutions. The participating institutions are the American Museum of Natural History, Astrophysical Institute Potsdam, University of Basel, University of Cambridge, Case Western Reserve University, University of Chicago, Drexel University, Fermilab, the Institute for Advanced Study, the Japan Participation Group, Johns Hopkins University, the Joint Institute for Nuclear Astrophysics, the Kavli Institute for Particle Astrophysics and Cosmology, the Korean Scientist Group, the Chinese Academy of Sciences (LAMOST), Los Alamos National Laboratory, the Max-Planck-Institute for Astronomy (MPIA), the Max-Planck-Institute for Astrophysics (MPA), New Mexico State University, Ohio State University, University of Pittsburgh, University of Portsmouth, Princeton University, the United States Naval Observatory, and the University of Washington. 


\section*{Data availability}
This study is based on publicly available data associated with Sloan Digital Sky Survey products. The description and sources of the catalogues used in this work are described in Section \ref{sec:data}.



\bibliographystyle{mnras}
\bibliography{bibtex} 



\appendix

\appendix

\section{Sensitivity of the Bar Likelihood to the Treatment of S0 Galaxies}
\label{app:s0_sensitivity}

To assess the sensitivity of our results to the treatment of
lenticular (S0) galaxies, we repeated the analysis of the bar likelihood
as a function of halo mass after separating the early-type population
according to the probabilistic S0 classification. Galaxies with
$P(\mathrm{S0})>0.80$ were identified as lenticular systems. The early-type
population was considered in three cases: (i) the complete early-type
sample, including S0 galaxies; (ii) the early-type population after
excluding the S0 population (early non-S0); and (iii) the S0 subsample
alone.

\begin{figure}
\centering
\includegraphics[width=\columnwidth]{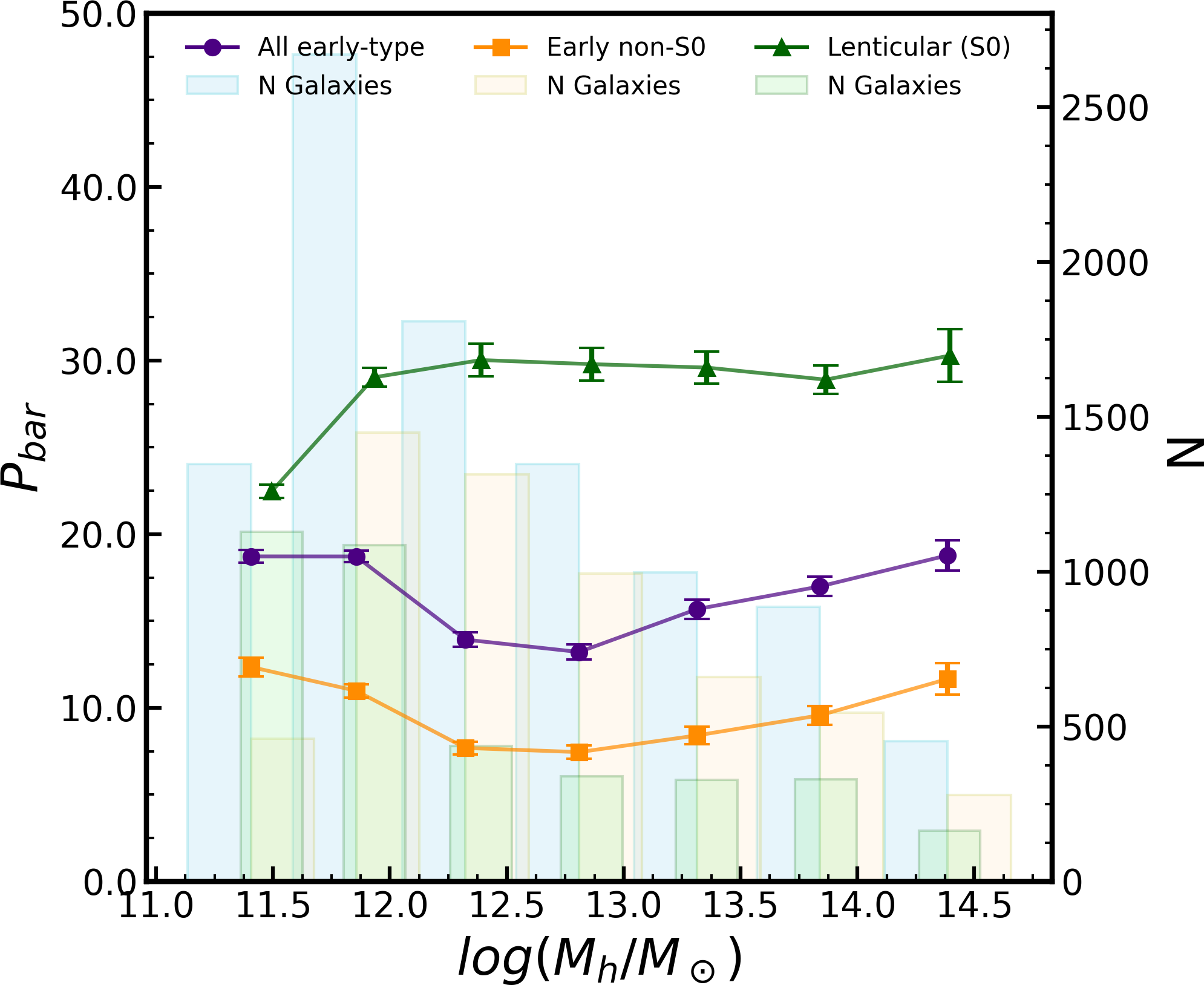}
\caption{Sensitivity test for the treatment of lenticular (S0) galaxies.
The bar likelihood, $P_{\rm bar}$, is shown as a function of halo mass for
the complete early-type population, the early-type population after
excluding galaxies with $P(\mathrm{S0})>0.80$, and the S0 subsample defined
by $P(\mathrm{S0})>0.80$. The exclusion of S0 galaxies changes the overall
level of $P_{\rm bar}$ but preserves the main qualitative trend with halo
mass, indicating that the latter is not driven by the inclusion of
lenticular galaxies.}
\label{fig:s0_sensitivity}
\end{figure}

Figure~\ref{fig:s0_sensitivity} compares the resulting $P_{\rm bar}$
distributions across the seven halo-mass bins. Excluding the S0 galaxies
produces a lower overall bar likelihood compared with the complete
early-type population, while the S0 subsample exhibits a higher bar
likelihood. This difference indicates that lenticular galaxies contribute
measurably to the absolute level of the bar likelihood within the
early-type population.

Importantly, however, the removal of the S0 population does not
substantially modify the overall shape of the $P_{\rm bar}$--$M_{\rm halo}$
relation. The main qualitative trend across the halo-mass bins is preserved
when S0 galaxies are excluded. Therefore, the dependence of the bar
likelihood on halo mass discussed in the main text is not driven by the
inclusion of lenticular galaxies in the early-type sample.

This test supports the robustness of our main result while also
demonstrating that the inclusion of S0 galaxies affects the normalization
of the bar likelihood. We therefore retain S0 galaxies within the early-type
population for the main morphological classification, while treating them
explicitly as a separate population when evaluating the sensitivity of the
results to their inclusion.

\section{Details of the residual environmental analysis}
\label{app:residual_environment}

The reference bar likelihood used in the residual analysis of
Section~\ref{subsec:residual_environment} is constructed from the satellite
population by combining the four morphological classes with four stellar-mass
bins, resulting in 16 morphology--stellar-mass reference cells. Table~\ref{tab:expected_pbar}
lists the number of satellite galaxies and the mean bar likelihood adopted
for each cell. These values provide the reference expectations used to
construct the environmental residuals shown in Fig.~\ref{fig:residual_environment}.

\begin{table}
\centering
\caption{Expected bar likelihood used to construct the environmental
residuals. The satellite reference sample is divided into four
morphological classes and four stellar-mass bins, resulting in 16
morphology--stellar-mass cells. $N$ is the number of satellite galaxies
in each cell, and $P_{\rm bar,exp}$ is the mean bar likelihood in that
cell}
\label{tab:expected_pbar}
\begin{tabular}{lcccc}
\hline
Morphology & Stellar-mass bin & $N$ & $P_{\rm bar,exp}$ \\
\hline
S0 & 9.021--9.631 & 25  & 0.1713 \\
S0 & 9.631--10.241 & 778 & 0.2545 \\
S0 & 10.241--10.850 & 424 & 0.3394 \\
S0 & 10.850--11.460 & 17 & 0.3253 \\
\hline
Early non-S0 & 9.021--9.631 & 11 & 0.0677 \\
Early non-S0 & 9.631--10.241 & 553 & 0.1314 \\
Early non-S0 & 10.241--10.850 & 892 & 0.1077 \\
Early non-S0 & 10.850--11.460 & 143 & 0.0594 \\
\hline
Late-I & 9.021--9.631 & 88 & 0.2993 \\
Late-I & 9.631--10.241 & 575 & 0.3379 \\
Late-I & 10.241--10.850 & 305 & 0.4364 \\
Late-I & 10.850--11.460 & 22 & 0.5161 \\
\hline
Late-II & 9.021--9.631 & 436 & 0.4178 \\
Late-II & 9.631--10.241 & 794 & 0.4378 \\
Late-II & 10.241--10.850 & 268 & 0.5198 \\
Late-II & 10.850--11.460 & 35 & 0.7198 \\
\hline
\end{tabular}
\end{table}

\bsp	
\label{lastpage}
\end{document}